\documentclass[final,5p,times,twocolumn]{elsarticle}
\usepackage{amssymb}
\usepackage{amsmath}
\usepackage{xcolor}
\usepackage{physics}
\usepackage{multirow}
\usepackage[normalem]{ulem}
\usepackage[percent]{overpic}
\usepackage{float}
\usepackage{cuted}
\usepackage{tikz-feynman}
\tikzfeynmanset{compat=1.1.0} 
\usepackage[colorlinks=true]{hyperref}
\newcommand{\orcid}[1]{\href{https://orcid.org/#1}{\includegraphics[width=8pt]{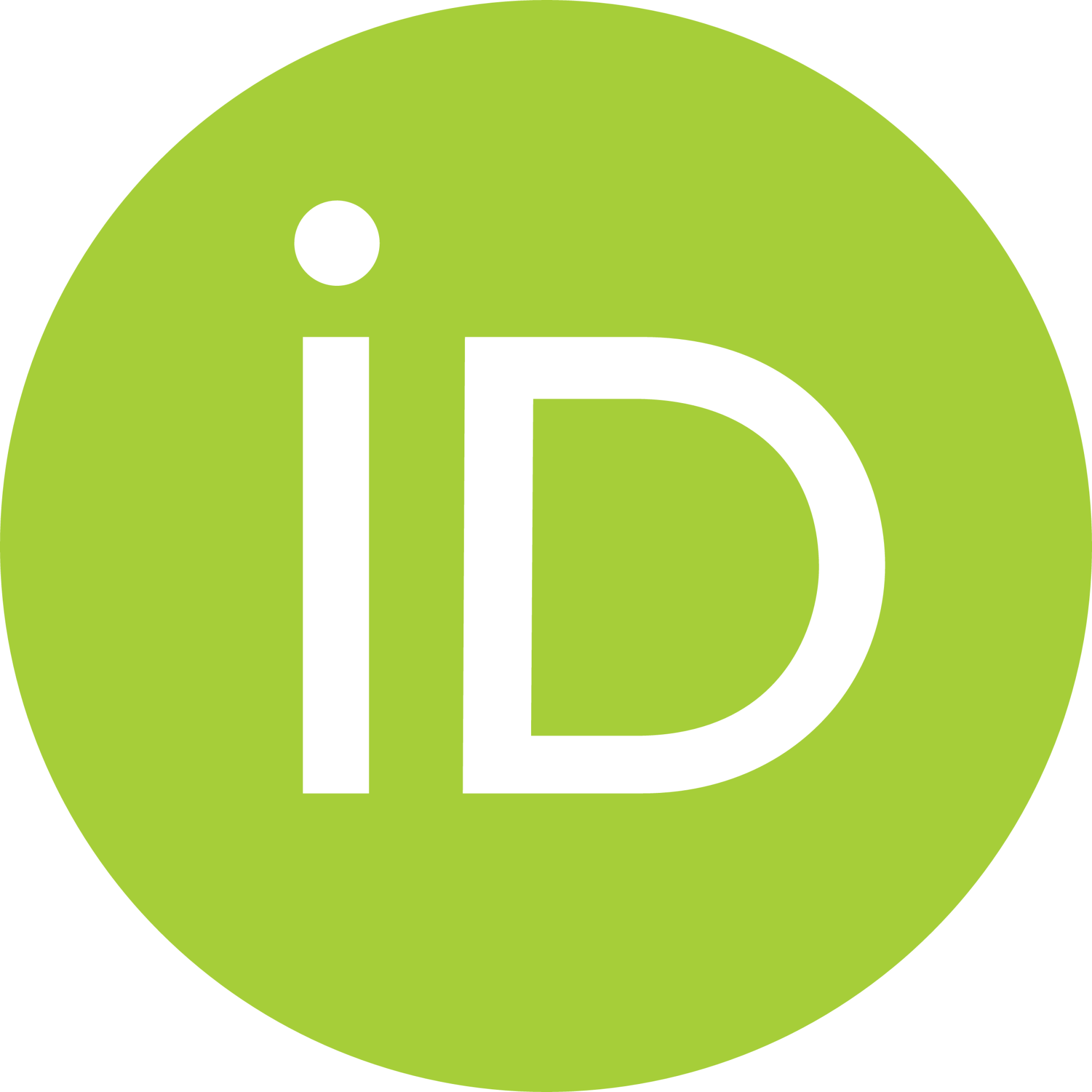}}}
\begin{document}
\begin{frontmatter}
\title{Distribution Functions of Radially Excited Pion using the Light-Front Quark Model}

\author[label1]{Ashutosh Dwibedi\corref{cor1}\orcid{0009-0004-1568-2806}}
\ead{ashutoshd@iitbhilai.ac.in}
\author[label2]{Satyajit Puhan\orcid{0009-0004-9766-5005}}
\ead{puhansatyajit@gmail.com}
\author[label1]{Sabyasachi Ghosh\orcid{0000-0003-1212-824X}}
\ead{sabya@iitbhilai.ac.in}
\author[label2]{Harleen Dahiya\orcid{0000-0002-3288-2250}}
\ead{dahiyah@nitj.ac.in}

\affiliation[label1]{organization={Department of Physics, Indian Institute of Technology Bhilai},
            addressline={Kutelabhata}, 
            city={Durg},
            postcode={491002}, 
            state={Chhattisgarh},
            country={India}}
\affiliation[label2]{organization={Computational High Energy Physics, Department of Physics, Dr. B R Ambedkar National Institute of Technology, Punjab, India.},
	addressline={Jalandhar},
	postcode={144008}, 
	state={Punjab},
	country={India}}

\begin{abstract}
We investigate the internal structure of the ground ($1S$) and the first two radially excited ($2S,3S$) states of the pion within the light-front quark model. The valence Fock sector is described using pure harmonic-oscillator eigenstates and mixed states formed as orthogonal linear combinations of these eigenfunctions. The optimal wavefunction parameters are determined through a variational procedure based on a QCD-motivated effective Hamiltonian. Using the resulting light-front wavefunctions, we study the pion distribution amplitude, parton distribution function, and electromagnetic form factor. After QCD evolution, the ground-state distribution amplitude and parton distribution function are found to be in good agreement with available experimental data. At the model scale, the parton distribution functions of the $1S$ and $2S$ states show clear sensitivity to state mixing, while the distribution amplitudes and electromagnetic form factors are weakly sensitive. In contrast, for the $3S$ state, all three observables exhibit a pronounced sensitivity to mixing. The decay constants of the mixed states are also found to decrease sequentially with increasing radial excitation.
\end{abstract}

\end{frontmatter}

\section{Introduction}
Understanding the complex internal structure of hadrons from first principles remains a challenging task in non-perturbative theoretical calculations. As a result, effective field theories and lattice quantum chromodynamics (QCD) have become essential tools for this purpose. So, from all perspectives, considerable effort is being devoted to studying the valence quark, gluon, and sea quark distributions in hadrons. The pion, being the lightest pseudoscalar meson, is an ideal hadron to study the internal structure due to its lower complexity than the baryons. A lot of progress has been made in understanding the 
ground-state distribution amplitude (DA), parton distribution function (PDF), and electromagnetic form factors (EMFF) theoretically \cite{Puhan:2024jaw,Roberts:2021nhw,Horn:2016rip,Agaev:2008zz,Lepage:1980fj,Puhan:2025pfs,Chai:2025xuz,Lan:2019vui,Venturini:2024nmh,Pasquini:2023aaf,Wu:2022iiu}. These distributions have been studied through lattice QCD in Refs. \cite{Cloet:2013tta,Zhang:2017bzy,Segovia:2013eca,JeffersonLabAngularMomentumJAM:2022aix,Lin:2025hka,Gao:2021xsm} for the case of pion. However, there is still a lack of information about these distribution functions from the experimental point of view. The upcoming electron-ion collider (EIC)~\cite{Accardi:2012qut} will provide more insight about the pion PDFs, EMFFs and three-dimensional structure through the Sullivan process~\cite{Hatta:2025ryj}.
\par Recent efforts have shifted toward probing the partonic structure of radially excited pions~\cite{Ridwan:2025fgs,Syahbana:2024hkc}. Predictions for the DA and PDF of the first excitation using Bethe–Salpeter amplitudes appear in Refs.~\cite{Wang:2025wrx,Xu:2025cyj}, while Ref.~\cite{Miramontes:2024fgo} examines its EMFF using a combined Schwinger–Dyson/Bethe–Salpeter framework. Light-front holographic QCD has been used to study mass spectra and decay constants of excited pion states~\cite{Ahmady:2022dfv}, and lattice QCD has recently provided first-principles estimates of their PDFs~\cite{Gao:2021hvs}. Besides pions,
the decay constants of excited mesons have also been explored within Bethe–Salpeter approaches~\cite{Holl:2004fr,Li:2016dzv} and lattice QCD~\cite{McNeile:2006qy,Mastropas:2014fsa}. Among effective approaches to hadron structure, the light-front constituent quark model (LFQM)~\cite{Dziembowski:1986dr,Cardarelli:1994ix,Choi:1997iq} has proven particularly successful due to its predictive capability for observables such as decay constants~\cite{Choi:2007se}, charge radii~\cite{Hwang:2001th}, and mass radii~\cite{Choi:2025rto}. Within this framework, a wide range of generalized quark correlation functions~\cite{Diehl:2003ny,Lorce:2025aqp} can be computed, providing direct links to hadronic scattering and decay observables. When mapped to impact-parameter space, these correlations yield spatial tomography of hadrons. In the LFQM approaches the light-front wavefunctions encode the hadron properties and ultimately connect us with the experimental results. The LFQM approach employed here follows the same principles as that of \cite{Choi:2015ywa}, where the space part of the front-form wavefunctions is determined through a variational analysis from the QCD-motivated effective Hamiltonian, while the spin part is obtained by the Melosh-Wigner transformation from the instant-form spin states. 
\par In this work, we investigate the ground state and the first two radially excited states of the pion within the framework of light-front quantization. In this approach, a meson state is expanded in Fock components, where the leading contribution arises from the valence quark–antiquark sector and higher components contain sea quarks and gluons. In the present analysis, the Fock-space expansion is truncated to the valence quark-antiquark sector only. For a judicious choice of model parameters, we carry out a variational analysis using harmonic oscillator eigenfunctions and their linear combinations, referred to as mixed wavefunctions. The analysis is based on a QCD-motivated effective Hamiltonian that includes kinetic, confining, Coulomb, and hyperfine interaction terms~\cite{Choi:2015ywa}. The harmonic parameter $\beta$ governs the shape of the light-front wavefunctions and plays a crucial role in determining mesonic properties. The hyperfine interaction, which is responsible for the mass splitting between pseudoscalar and vector mesons with identical quark content, is treated non-perturbatively. This is achieved by smearing the delta-function interaction with a finite-width Gaussian following Ref.~\cite{Choi:2015ywa}. Such a treatment is justified by the large mass splittings observed in the light meson sector. In this work, we employ a standard contour plot method, rather than the hit-and-trial method, to calculate the parameters.
%We are also trying to calculate the parameters using the machine learning method in the near future.
Once the optimal harmonic parameters and mixing angles are determined from the variational procedure, the corresponding light-front wavefunctions are used to systematically study the internal structure of the pion. In particular, we investigate the DA, PDF, and EMFF for both pure and mixed radially excited states. The ground state DAs and PDFs have been evolved from the model scale to higher momentum scales using the Efremov-Radyushkin-Brodsky-Lepage (ERBL) \cite{Brodsky:1981jv} and Dokshitzer–Gribov–Lipatov–Altarelli–Paris (DGLAP) evolution equations \cite{Miyama:1995bd,Hirai:1997gb,Hirai:1997mm}, respectively, to compare with existing experimental data and theoretical predictions. In addition, the decay constants and charge radii of the excited pion states are also evaluated using the corresponding light-front wavefunctions and have been compared with available theoretical models.

The paper is organized as follows. In Sec.~\eqref{framew}, we present the formalism and the essential ingredients for constructing the light-front wavefunctions of the ground and radially excited pion states, including both pure and mixed configurations. In Sec.~\eqref{result}, these wavefunctions are employed to define the pion DA, PDF, and EMFF, which are then computed for both pure and mixed state wavefunctions. Finally, Sec.~\eqref{sum} contains a summary of our results and concluding remarks.

\section{Framework and model descriptions}\label{framew}
The QCD-motivated effective Hamiltonian in the rest frame of a meson can be expressed~\cite{Choi:1997iq,Godfrey:1985xj},
\begin{equation}
   H_{q_{1}\bar{q}_{2}}= \sqrt{m_{1}^{2}+\vec{p}^{2}} + \sqrt{m_{2}^{2}+\vec{p}^{2}} +V_{q_{1}\bar{q}_{2}},\label{AD1}
\end{equation}
where the first two terms in the right hand side are the free parts of the relativistic kinetic energy for the constituent quark $q_{1}$ (with mass $m_1$) and antiquark $\bar{q}_{2}$ (with mass $m_2$)\footnote{ In the rest frame of meson, the constituent quarks move with equal and opposite momentum, i.e., $\vec{p}_{1}=-\vec{p}_{2}=\vec{p}$.}. The total effective potential of the quark-antiquark system takes the form $V_{q_{1}\bar{q}_{2}}=V_{Con} + V_{Cou} + V_{Hyp}$, where $V_{Con}$, $V_{Cou}$, and $V_{Hyp}$, are respectively, the confining, coulombic and hyperfine interaction potentials.  While in the literature~\cite{Choi:1997iq,Dhiman:2019ddr,Pandya:2024qoj} many confining potentials are in use we take the usual linear confining potential~\cite{Choi:2015ywa} $V_{Con}=a+br$. The effective one gluon exchange potential ($V_{Oge}$) between the quarks are captured by the coulomb $V_{Cou}=-\frac{4\alpha_{s}}{3r}$ ($\alpha_s$ is the strong coupling) and hyperfine $V_{Hyp}=\frac{2}{3}\frac{\vec{S}_{1}\cdot \vec{S}_{2}}{m_1m_2}~\nabla^{2}V_{Cou}$ interactions. It is important to point out that the spin-spin interaction contained in the $V_{Hyp}$ splits the energy levels of otherwise degenerate states of pseudoscalar and vector mesons of the same quark content, e.g., pion and rho meson. Following the prescriptions in Ref.~\cite{Choi:2015ywa}, we smear out the hyperfine interaction containing the delta function with a smearing parameter $\sigma$ as, $V_{Hyp}=\frac{32\pi\alpha_{s}}{9}\frac{\vec{S}_{1}\cdot \vec{S}_{2}}{m_1m_2}\delta^{3}(\vec{r})\rightarrow \frac{32\pi\alpha_{s}}{9}\frac{\vec{S}_{1}\cdot \vec{S}_{2}}{m_1m_2} \frac{\sigma^{3}}{\pi^{3/2}}e^{-\sigma^{2}r^{2}}$. We now seek the eigenvalues (masses), eigenfunctions (wavefunctions) of the Hamiltonian given in Eq. \eqref{AD1}. For a heavy quark system the hamiltonian becomes essentially non-relativistic with the free part of the energy simplifies to $\sqrt{m_{1}^{2}+\vec{p}^{2}}\approx m_{1}+\frac{\vec{p}^{2}}{2m_{1}}$ and  $\sqrt{m_{2}^{2}+\vec{p}^{2}}\approx m_{2}+\frac{\vec{p}^{2}}{2m_{2}}$. Additionally, the hyperfine interaction in heavy quark systems can be treated as a perturbation. Even in this case, it remains challenging to determine the exact analytical eigenvalues and eigenfunctions, and one often resorts to various approximation methods~\cite{Lucha:1991vn}. In the present work, we apply the variational principle to obtain the wavefunctions and masses of the first three low-lying radially excited states of the pion. We treat the full Hamiltonian in Eq. \eqref{AD1} nonperturbatively in the analysis, and use a linear combination of the first three harmonic oscillator wavefunctions, i.e., 
\begin{eqnarray}
&&\begin{pmatrix}
\varPhi_{1S} \\
\varPhi_{2S} \\
\varPhi_{3S}
\end{pmatrix}_{3\times 1}
= \mathcal {R}
\begin{pmatrix}
\varphi_{1S} \\
\varphi_{2S} \\
\varphi_{3S}
\end{pmatrix}_{3\times 1},\text{where}~\mathcal{R}=\begin{pmatrix}
\mathcal{R}_{1} \\
\mathcal{R}_{2}\\
\mathcal{R}_{3}
\end{pmatrix}_{3\times 3} \text{is the}\nonumber\\
&&\text{mixing matrix.}
\label{AD2}
\end{eqnarray}
The block matrices $(\mathcal{R}_{i})_{1 \times 3}$ defining the rows of the mixing matrix $\mathcal{R}_{3\times 3}$ is given by~\cite{Ridwan:2024ngc,Ridwan:2024hyh}
\begin{eqnarray}
\mathcal{R}_1&=&[\cos{\phi}\cos{\theta},~  \sin{\phi}\cos{\theta},~ \sin{\theta}],\nonumber\\
\mathcal{R}_2&=&[-\sin{\phi}\cos{\theta}-\cos{\phi}\sin^{2}{\theta},~\nonumber\\
&&\cos{\phi}\cos{\theta}-\sin{\phi}\sin^{2}{\theta},~ \sin{\theta}\cos{\theta}]\nonumber\\
\mathcal{R}_{3}&=&[-\sin{\phi}\cos{\theta}-\cos{\phi}\sin{\theta}\cos{\theta},~\nonumber\\
&&-\cos{\phi}\sin{\theta}-\sin{\phi}\sin{\theta}\cos{\theta},~\cos^{2}{\theta}],\nonumber
\end{eqnarray}
where $\theta$ and $\phi$ are the mixing parameters to be determined. The harmonic oscillator eigenfunctions are expressed as,
\begin{eqnarray}
    && \varphi_{1S}=\frac{1}{\pi^{3/4}\beta^{3/2}} e^{-\vec{p}^{2}/2\beta^{2}},\label{AD3}\\
    && \varphi_{2S}=\frac{2\vec{p}^{2}-3\beta^{2}}{\sqrt{6}\pi^{3/4}\beta^{7/2}} e^{-\vec{p}^{2}/2\beta^{2}},\label{AD4}\\
    && \varphi_{3S}=\frac{15\beta^{4}-20\vec{p}^{2}\beta^{2}+4\vec{p}^{4}}{2\sqrt{30}\pi^{3/4}\beta^{11/2}} e^{-\vec{p}^{2}/2\beta^{2}},\label{AD5}
\end{eqnarray}
where $\beta$ is the harmonic parameter and $\vec{p}^{2}=\vec{p}_{\perp}^{2}+p_{z}^{2}$.

\begin{figure}[htbp]
        \centering
        \includegraphics[width=0.47\textwidth]{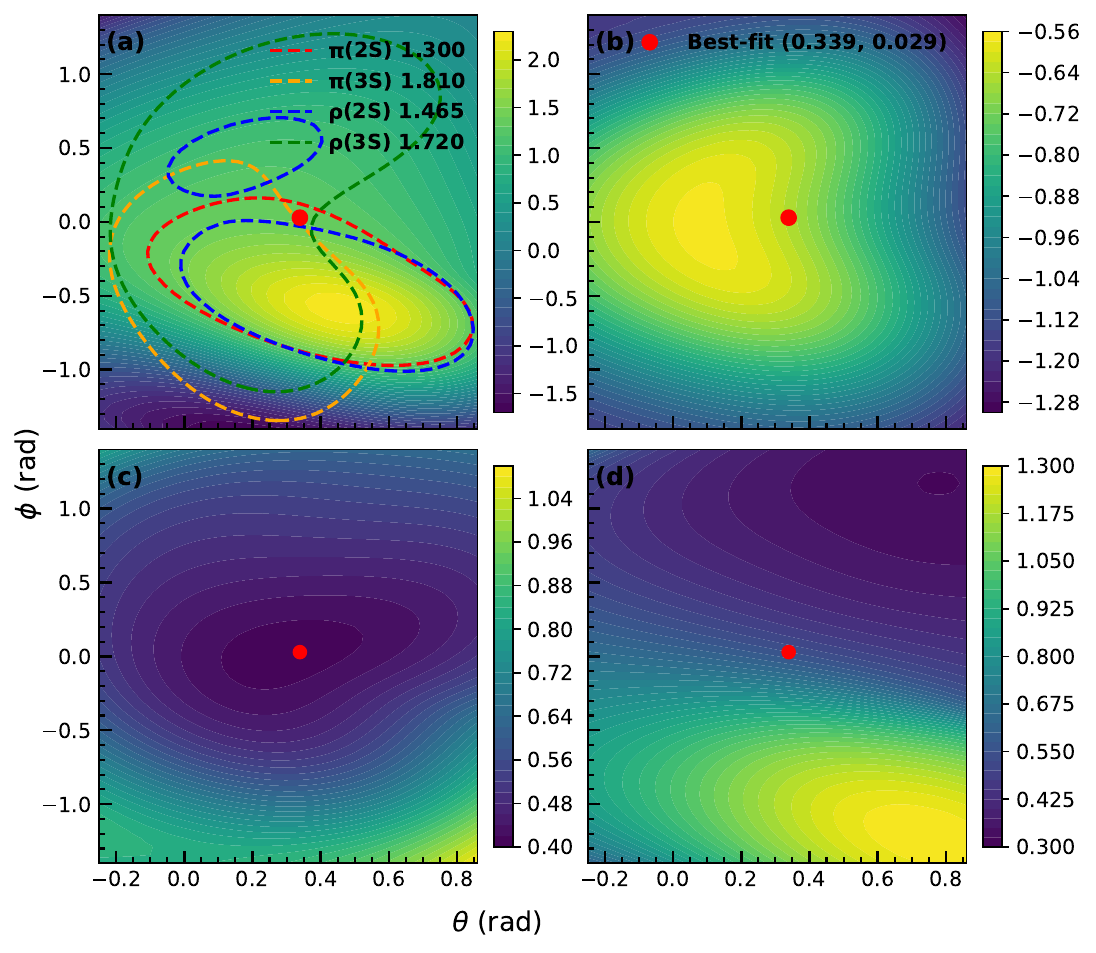}
        \caption{(Color online) (a) Experimental mass contours of $\pi(3S)$, $\rho(2S)$, and $\rho(3S)$ overlaid on the $\pi(2S)$ base surface; (b) variation of the potential parameter $a$ with the mixing angles; (c) variation of the potential parameter $\alpha_s$ with the mixing angles; and (d) variation of the harmonic parameter $\beta_{\pi}$ with the mixing angles. The best-fit point is indicated by a red dot in all contour plots.}
        \label{fig:cont}
    \end{figure}

Our ultimate goal in this section is to fit the excited state masses of the pion by varying our model parameters. Before delving into the details of our variational method, let us count the number of parameters of our model. We have a total of nine parameters to be fixed: potential parameters ($a$, $b$, $\alpha_s$, and $\sigma$), light-quark mass ($m_{u}=m_{d}=m$), mixing angles ($\theta,\phi$), and the harmonic parameters ( $\beta_{\pi},\beta_{\rho}$). Applying the variational principle for $\pi$ and $\rho$ mesons by treating $\beta_{\pi}$ and $\beta_{\rho}$ as variational parameters we get,
\begin{eqnarray}
  && \frac{\partial \bra{\varPhi^{\pi}_{1S}} H_{u\bar{d}}\ket{\varPhi^{\pi}_{1S}} }{\partial \beta_{\pi} }=\frac{\partial \bra{\varPhi^{\rho}_{1S}} H_{u\bar{d}}\ket{\varPhi^{\rho}_{1S}} }{\partial \beta_{\rho} }=0. \label{AD6} 
\end{eqnarray}
The total wavefunctions are given by a product of space and spin part, $\varPhi^{\pi}_{1S}=\varPhi_{1S}(\beta_{\pi})~\chi_{s}$ and $\varPhi^{\rho}_{1S}=\varPhi_{1S}(\beta_{\rho})~\chi_{t}$ where $\chi_{s}$ and $\chi_{t}$ are respectively the spin singlet \cite{Arifi:2022pal} and triplets \cite{Tanisha:2025qda}. To this end, we made some educated guesses about fixing some of the model parameters from the existing literature~\cite{Choi:2015ywa}. We take the constituent quark mass $m=0.205$ GeV, the string tension $b=0.18$ GeV$^{-2}$, and the smearing parameter $\sigma=0.423$ GeV. Now we impose two more constraints to reduce our number of free parameters,
\begin{eqnarray}
    && \bra{\varPhi^{\pi}_{1S}} H_{u\bar{d}}\ket{\varPhi^{\pi}_{1S}}= M^{\pi}_{1S}, \bra{\varPhi^{\rho}_{1S}} H_{u\bar{d}}\ket{\varPhi^{\rho}_{1S}}= M^{\rho}_{1S}.\label{AD7}
\end{eqnarray}
Eqs.~\eqref{AD6} and \eqref{AD7} can be solved simultaneously to obtain $a$, $\alpha_s$, $\beta_{\pi}$, and $\beta_{\rho}$ as functions of the free parameters $\theta$ and $\phi$. The masses can now be obtained by taking the expectation value of the Hamiltonian provided in Eq. \eqref{AD1} with the use of $a (\theta, \phi)$, $\alpha_s(\theta,\phi)$, $\beta_{\pi}(\theta,\phi)$, and $\beta_{\rho}(\theta,\phi)$. For finding out the values of ($\theta$,$\phi$) that best fit\footnote{Practically, we minimized the quantity $\sqrt{\sum_{i}(M_{i}(\theta,\phi)-M_{i}^{\rm ex})^{2}}$, where $i$ runs over the excited states.} the experimental masses of $\pi(2S)$ and $\pi(3S)$, we take the help of a contour plot. Figure~\eqref{fig:cont} (a) represents the variation of the mass of the state $\pi(2S)$ against the mixing angles $\theta$ and $\phi$. It also displays the exact experimental mass contour of $\pi(2S)$ (red dashed line) along with other excited states $\pi(3S)$ (yellow dashed line), $\rho(2S)$ (blue dashed line), and $\rho(3S)$ (green dashed line). 
From this analysis, we find the best-fit values of the $\theta=0.339$ rad$=19.42^{\circ}$ and $\phi=0.029$ rad $=1.64^{\circ}$. The corresponding potential parameter $a=-0.647$ GeV and strong coupling constant $\alpha_{s}=0.410$. The masses and the harmonic parameters obtained with the best-fit values of the mixing angles are provided in  Table~\eqref{T1}. In Table~\eqref{T1}, we also give the corresponding values of the parameters in the no mixing case, i.e., $\theta=\phi=0$. The value of $\chi^{2}=\sum_{i}\frac{(M_{i}-M_{i}^{\rm ex})^{2}}{M_{i}^{\rm ex}}$ for the mixing and no-mixing (or, pure) case is obtained as $6.51$ MeV and $84.153$ MeV, respectively. One can observe that the incorporation of mixing angles reduces the $\chi^{2}$ and gives a better fit to the PDG masses. This implies that one can reduce the $\chi^{2}$ by including more mixing angles (greater than two) in the calculation. 
\begin{table}[H]
\centering
\begin{tabular}{|c|c|c|c|c|}
\hline
 Meson & $\pi(2S)$ & $\pi(3S)$ & $\rho(2S)$ & $\rho(3S)$ \\ \hline
 $a,\alpha_{s}$ (Pure) & \multicolumn{4}{c|}{-0.580,~0.438}  \\ \hline
$a,\alpha_{s}$ (Mixing) & \multicolumn{4}{c|}{-0.647,~0.41}  \\ \hline
$\beta$ (Pure) & \multicolumn{2}{c|}{0.634} & \multicolumn{2}{c|}{0.332} \\ \hline
$\beta$ (Mixing) & \multicolumn{2}{c|}{0.550} & \multicolumn{2}{c|}{0.328} \\ \hline
Mass (Pure) & 1.331 & 2.120 & 1.442 & 1.947 \\ \hline
Mass (Mixing)& 1.365 & 1.783 & 1.430 & 1.780 \\\hline
Mass (PDG) \cite{10.1093/ptep/ptac097}             & 1.300 & 1.810 & 1.465 & 1.720 \\[1ex]
\hline
\end{tabular}
\caption{Comparison of calculated and PDG masses \cite{10.1093/ptep/ptac097} of excited $\pi$ and $\rho$ mesons with and without mixing. The corresponding potential and harmonic parameters are also listed. All quantities are in GeV except for $\alpha_s$.}
\label{T1}
\end{table}
We should stress that the $\beta$ in the pseudo scalar and vector sectors becomes different in our analysis. This is a result of treating hyperfine interaction non-perturbatively~\cite{Choi:2015ywa}. We end the discussion by emphasizing that the above method can, in general, be applied to know the mass spectra of other mesons containing $s,c,$ and $b$ quarks.

Now, with the instant-from wavefunctions of the pion at our hands, we shift our attention to describing the pion states at a given light-front time $z^{+}=t+z=0$. Light-front descriptions are increasingly used in hadron structure phenomenology because they incorporate the relativistic nature of the partons, and the hadronic matrix elements take particularly simple forms. After briefly addressing the Fock state description of the pion at $z^{+}=t+z=0$, we would move on to connect the front-form wavefunctions with the instant-form wavefunctions obtained earlier in Eqs.~\eqref{AD2} to \eqref{AD5}. Since all these discussions would be valid for the ground as well as the excited state, we temporarily omit the state labels $1S,2S$, and $3S$, and recover them whenever necessary. In LFQM the state of the pion (both ground and excited states) can be labeled by eigen values of six mutually commuting operators: Mass $\hat{M}^{2}=\hat{P}^{\mu}\hat{P}_{\mu}$,  three space-like momenta $\hat{P}^{+}$, $\hat{\vec{P}}_{\perp}$, and squared Pauli-Lubansky Vector $S^{2}\equiv\hat{W}^{\mu}\hat{W}_{\mu}$ along with its longitudinal component $\hat{W}^{+}\equiv \hat{S}_{z}$. The Fock state expansion of a pion can be expressed as a combination of constituents (quark, antiquark, gluons, and sea quark) as \cite{Pasquini:2023aaf}
\begin{equation}
\ket{\Pi (P^+,\vec{P}_{\perp})}=\ket{q\bar{q}}+\ket{q\bar{q}g}+\ket{q\bar{q}q\bar{q}}+\dots,\label{AD8}
\end{equation}
where the leading term in the RHS of Eq. \eqref{AD8} is called the valence Fock sector, followed by states having gluons and sea quarks. In writing the state of the pion in the LHS of Eq. \eqref{AD8}, we ignore the obvious eigen labels $S_{z}=0$, $S=0$, and the mass $M$. The Fock space description of the pion can be saturated by only considering the leading (valence) Fock sector constructed from the constituent $u$ and $\bar{d}$ quarks $\ket{u\bar{d} (x_{i}P^{+},\vec{k}_{\perp i}+x_{i}\vec{P}_{\perp},\lambda_{i})}$, where $x_{i}=\frac{k_{i}^{+}}{P^{+}}$, $\vec{k}_{\perp i}$ and $\lambda_{i}$ are the boost-invariant longitudinal momentum fraction, transverse (internal) momentum, and helicity carried by the $i$th constituent parton, respectively. These variables obey the following relation because of the momentum conservation, $\sum_{i}x_{i}=1$ and $\sum_{i}\vec{k}_{\perp i}=0$. This approach of representing the pion is customarily known as the light-front constituent quark model in the literature. Keeping only the valence Fock sector, the state of the pion can be written as,

\begin{eqnarray}
|\Pi (P^+,\vec{P}_{\perp})\rangle &=&\sum_{\lambda_{1},\lambda_{2}}\int \frac{dx_{1}dx_{2}d^{2}{\bf k}_{\perp 1}d^{2}{\bf k}_{\perp 2}}{2(2\pi)^{3}~2(2\pi)^{3}\sqrt{x_{1}x_{2}}}~2(2\pi)^{3}\nonumber\\
&&\delta(1-x_{1}-x_{2})~\delta^{2}(\vec{k}_{\perp 1}+\vec{k}_{\perp 2})\nonumber\\
&&\varPsi^{\pi}(x_{1},x_{2},{\vec k}_{\perp 1},{\vec k}_{\perp 2},\lambda_{1},\lambda_{2})\nonumber\\
&&|x_{1}P^{+},x_{1}\vec{P}_{\perp}+{\vec k}_{\perp 1},\lambda_{1}\rangle\nonumber\\
&&|x_{2}P^{+},x_{2}\vec{P}_{\perp}+{\vec k}_{\perp 2},\lambda_{2}\rangle\nonumber\\
&=&\int \frac{dx \, d^2 \vec{k}_{\perp}}{  16 \pi^3 \sqrt{x(1-x)}}\nonumber\\
&&\big[\varPsi^{\pi} (x,\vec{k}_{\perp},\uparrow,\uparrow)|x P^+, \vec{k}_{\perp}, \uparrow, \uparrow \rangle  \,\nonumber\\
&& + \varPsi^{\pi} (x,\vec{k}_{\perp},\uparrow,\downarrow) \, |x P^+, \vec{k}_{\perp}, \uparrow, \downarrow \rangle\nonumber\\
&&+ \varPsi^{\pi} (x,\vec{k}_{\perp},\downarrow,\uparrow) \, |x P^+, \vec{k}_{\perp},  \downarrow,\uparrow \rangle\nonumber\\
&&+ \varPsi^{\pi} (x,\vec{k}_{\perp},\downarrow,\downarrow) \, |x P^+, \vec{k}_{\perp}, \downarrow, \downarrow \rangle \big] \,\label{AD9}, 
\end{eqnarray} 
  where to obtain the last line, we integrate over the coordinates of the antiquark ($x_{2}$ and $\vec{k}_{\perp 2}$), omit the antiquark momentum labels in the two particle Fock state, and ignore the subscripts in the quark variables for simplicity. The $\varPsi^{\pi}(x,\vec{k}_{\perp},\lambda_{1},\lambda_{2})$ appearing in Eq. \eqref{AD9} are the light-front wavefunctions of the pion. To this end we connect the instant-form wavefunctions $\varPhi^{\pi}(\vec{p},s_{1},s_{2})=\varPhi(\vec{p})~\chi_{s}(s_{1},s_{2})$ obtained in the CM frame of the quark-antiquark system with the front-form wavefunction $\varPsi^{\pi}(x,\vec{k}_{\perp},\lambda_{1},\lambda_{2})=\psi(x,\vec{k}_{\perp})\mathcal{S}(x,\vec{k}_{\perp},\lambda_{1},\lambda_{2})$. Usually, one links the instant-form momentum ($\vec{p}$) with front-form ($x,\vec{k}_{\perp}$) by following the Brodsky-Huang-Lepage prescription--$\vec{p}_{\perp}=\vec{k}_{\perp}$ and $p_{z}=(x-\frac{1}{2})M_{0}$, where $M_{0}^{2}=\frac{\vec{k}_{\perp}^{2}+m^{2}}{x(1-x)}$ is the invariant mass \cite{Lepage:1980fj}. As a result we have\footnote{ We are using $\varPhi$ to denote mixed wavefunctions which reduce to pure eigen wavefunctions $\varphi$ defined in Eqs.~\eqref{AD3} to \eqref{AD5} when $\theta=\phi=0.$} $\psi(x,\vec{k}_{\perp})=\sqrt{2(2\pi)^{3}}\sqrt{\frac{\partial p_{z}}{\partial x}}\varPhi(\vec{p}^{2}=\frac{\vec{k}_{\perp}^{2}+m^{2}}{4x(1-x)}-m^{2})$, where $\sqrt{\frac{\partial p_{z}}{\partial x}}=\sqrt{\frac{M_{0}}{4x(1-x)}}$. The instant-form spin states $\chi_{s}(s_{1},s_{2})$
%$=\frac{1}{\sqrt{2}}\left(\ket{1/2,\uparrow}^{1}_{T}\ket{1/2,\downarrow}^{2}_{T}-\ket{1/2,\downarrow}^{1}_{T}\ket{1/2,\uparrow}^{2}_{T}\right)$
are transformed to the front-form spin states $\mathcal{S}(x,\vec{k}_{\perp},\lambda_{1},\lambda_{2})$ $via$ the interaction independent Melosh-Wigner transformation, $\ket{J, \lambda}_{F}=\sum_{s} U^{J}_{s\lambda}\ket{J ,s}_{T}$, where $U^{J}_{s\lambda}$ is the unitary operator connecting the instant-form basis $\ket{}_{T}$ to front-form basis $\ket{}_{F}$. Equivalently, the spin wavefunctions can also be obtained by choosing an appropriate quark-meson vertex in terms of light-front spinors, i.e., $\mathcal{S}=\frac{1}{\sqrt{2M_{0}}}\bar{u}(xP^{+},\vec{k}_{\perp},\lambda_{1})~\gamma_{5}~v((1-x)P^{+},-\vec{k}_{\perp},\lambda_{2})$,
\begin{eqnarray}
&&\mathcal{S}(\uparrow,\uparrow)=-\frac{k_{x}-ik_{y}}{\sqrt{2(m^{2}+\vec{k}_{\perp}^{2})}},~\mathcal{S}(\uparrow,\downarrow)=\frac{m}{\sqrt{2(m^{2}+\vec{k}_{\perp}^{2})}} , \nonumber\\ 
&&\mathcal{S}(\downarrow,\uparrow)=-\frac{m}{\sqrt{2(m^{ 2}+\vec{k}_{\perp}^{2})}},~\mathcal{S}(\downarrow,\downarrow)=-\frac{k_{x}+ik_{y}}{\sqrt{2(m^{ 2}+\vec{k}_{\perp}^{2})}},\nonumber \\
\label{AD10}
\end{eqnarray} 
where the momentum dependence is suppressed in the argument of $\mathcal{S}$ for simplicity. It is easy to see that the following normalization condition holds,
\begin{eqnarray}
&&\sum_{\lambda_{1}\lambda_{2}}\int \frac{dxd^{2}\vec{k}_{\perp}}{2(2\pi)^{3}}\lvert\varPsi^{\pi}(x,\vec{k}_{\perp},\lambda_{1},\lambda_{2})\rvert^{2}=1. \label{AD11}\end{eqnarray}
\section{Results}\label{result}
In this section, we use the description of the pion state provided in Eq.~\eqref{AD9} along with the light-front wavefunctions analyzed in Sec.~\eqref{framew} to describe the internal structure of the pion by computing its DA, PDF, and EMFF. By first obtaining the general expression of these quantities in terms of the wavefunctions, we compute DA, PDF and EMFF for two cases: with mixing corresponds to $\theta=19.42^{\circ}$ and $\phi=1.64^{\circ}$ and without mixing correspond to $\theta=\phi=0$.  
\subsection{Decay Constant and Distribution Amplitudes (DAs)}
The meson DAs are defined in terms of the matrix elements of non-local operators that are sandwiched between the vacuum and the meson states \cite{Hwang:2010hw,Chang:2016ouf,Ahmady:2016ufq}
\begin{eqnarray}
\langle 0|\bar{\vartheta}_{d}(0) \gamma^\mu \gamma_5 \vartheta_{u}(0)|\Pi (P) \rangle &=&  f_{m}P^{\mu}\int_{0}^{1} \phi_{\rm DA}(x)~ dx.\label{AD12}
\end{eqnarray}
Substituting the quark field operators $\bar{\vartheta}_{d}(0)$ and $\vartheta_{u}(0)$, the explicit forms of pion DA in our LFQM is given by \cite{Choi:2015ywa}
\begin{eqnarray}
\phi_{\rm DA}(x) &=& \frac{2\sqrt{6}}{f_m}\int \frac{d^2\vec{k}_{\perp}}{16 \pi^3} \psi(x,\vec{k}_{\perp}) \frac{m}{\sqrt{m^2+\vec{k}^2_{\perp}}}\nonumber\\
&=& \frac{2\sqrt{6}}{f_m}\int \frac{d^2\vec{k}_{\perp}}{\sqrt{16 \pi^3}}\sqrt{\frac{\partial p_{z}}{\partial x}} \varPhi(x,\vec{k}_{\perp}) \frac{m}{\sqrt{m^2+\vec{k}^2_{\perp}}}.\nonumber\\\label{AD13}
\end{eqnarray}
Substituting the value of the Jacobian $\sqrt{\frac{\partial p_{z}}{\partial x}}=\frac{1}{\sqrt{4x(1-x)}}\left(\frac{\vec{k}_{\perp}^{2}+m^{2}}{x(1-x)}\right)^{1/4}$ in Eq.~\eqref{AD13}, we have,
\begin{eqnarray}
\phi_{\rm DA}(x) &=& \frac{\sqrt{6}}{f_m}\int \frac{d^2\vec{k}_{\perp}}{\sqrt{16 \pi^3}} \frac{m ~\varPhi(x,\vec{k}_{\perp})}{(x(1-x))^{3/4} ~(m^{2}+\vec{k}_{\perp}^{2})^{1/4}}\nonumber~.\\ \label{AD14}
\end{eqnarray}
From Eq.~\eqref{AD14} we can obtain the decay constant of the pion imposing the normalization: $\int \phi_\pi(x) ~dx=1$.
%%%
\begin{figure}[H]
\centering
\begin{overpic}[width=0.48\textwidth]{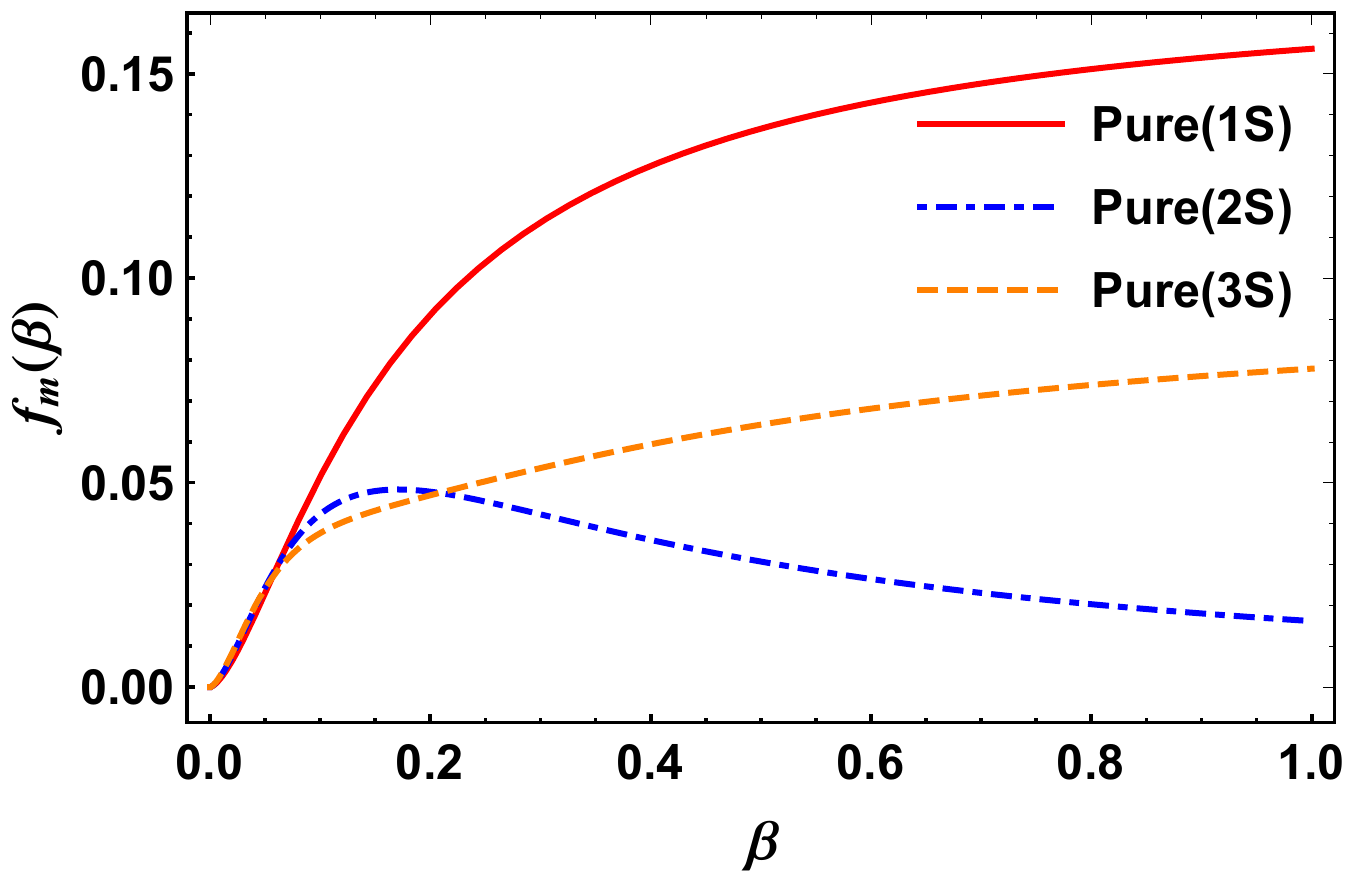}
    \put(17,58){\small (a)}
\end{overpic}
\hfill
\begin{overpic}[width=0.48\textwidth]{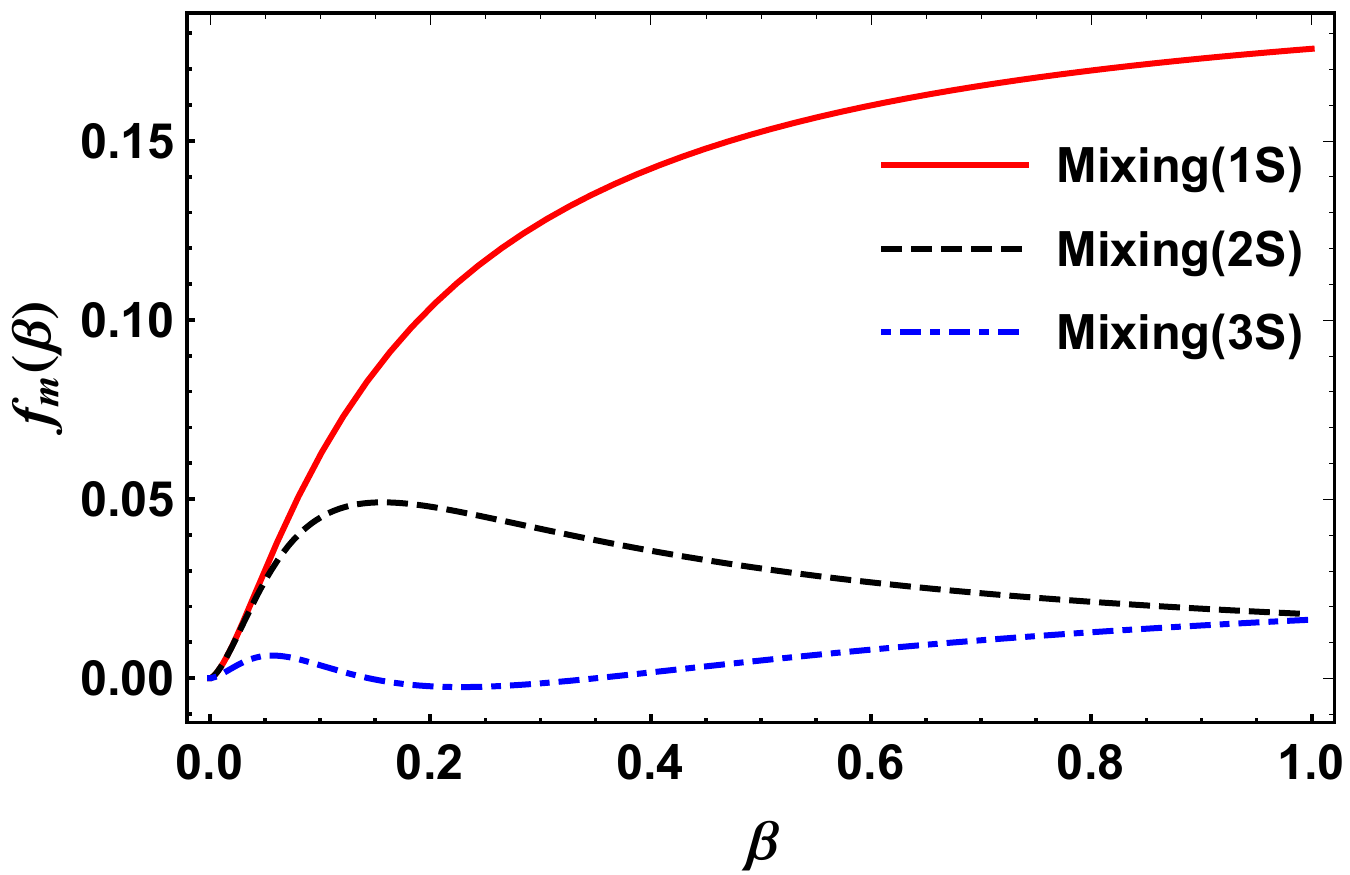}
    \put(17,58){\small (b)}
\end{overpic}
\caption{(Color online) Variation of Decay constant as a function of harmonic parameter $\beta$ for pure states (a) and for the case of mixing (b).}
\label{fig1}
\end{figure}
%%%%%%%%%%%%%%%%%%%%%%%%
\begin{table}[H]
\centering
\begin{tabular}{|c|c|c|c|}
\hline
 $f_m$ & $\pi(1S)$ &$\pi(2S)$ & $\pi(3S)$ \\ \hline
 This work (Pure) & 146.3 & 25.9 &  69.9 \\ \hline 
 This work (Mixing) & 156.4 & 28.6 & 6.5  \\ \hline 
Ref. \cite{Ahmady:2022dfv} & 166.5 & 1.44 &  0.65 \\ \hline 
Ref. \cite{Arndt:1999wx}   &  91.9 & 18.6 &  42.3 \\ \hline
Ref. \cite{Xu:2025cyj} & 98 & 8 & --  \\ \hline 
PDG \cite{10.1093/ptep/ptac097} & 130.2$\pm$ 1.7 & -- & -- \\[1ex]
\hline
\end{tabular}
\caption{Comparison of our calculated decay constant of the pion radially excited state with available theoretical predictions \cite{Ahmady:2022dfv,Arndt:1999wx,Xu:2025cyj} and PDG data \cite{10.1093/ptep/ptac097}}
\label{T2}
\end{table}
We start the discussion on the DA of the pions by first observing the behavior of the decay constants with the harmonic parameter $\beta$. In Fig.~\eqref{fig1}, we display the variation of the decay constant with the harmonic parameter $\beta$, first for the pure state Fig.~\eqref{fig1} (a) and then for the mixed states Fig.~\eqref{fig1} (b). We observe the decay constant for the ground state pions to monotonically increase with $\beta$, for both pure and mixed states. For the mixed $1S$ state decay constant increases with a slightly faster rate, reaching a value $\approx 175$ MeV for $\beta=1$ GeV, whereas the same for the pure state is about $155$ MeV. For the excited states, one should expect the decay constant to be significantly less than the ground state, as advocated by~\cite{Arifi:2022pal}.
%%%%%
\begin{figure}[H]
\centering
\begin{overpic}[width=0.48\textwidth]{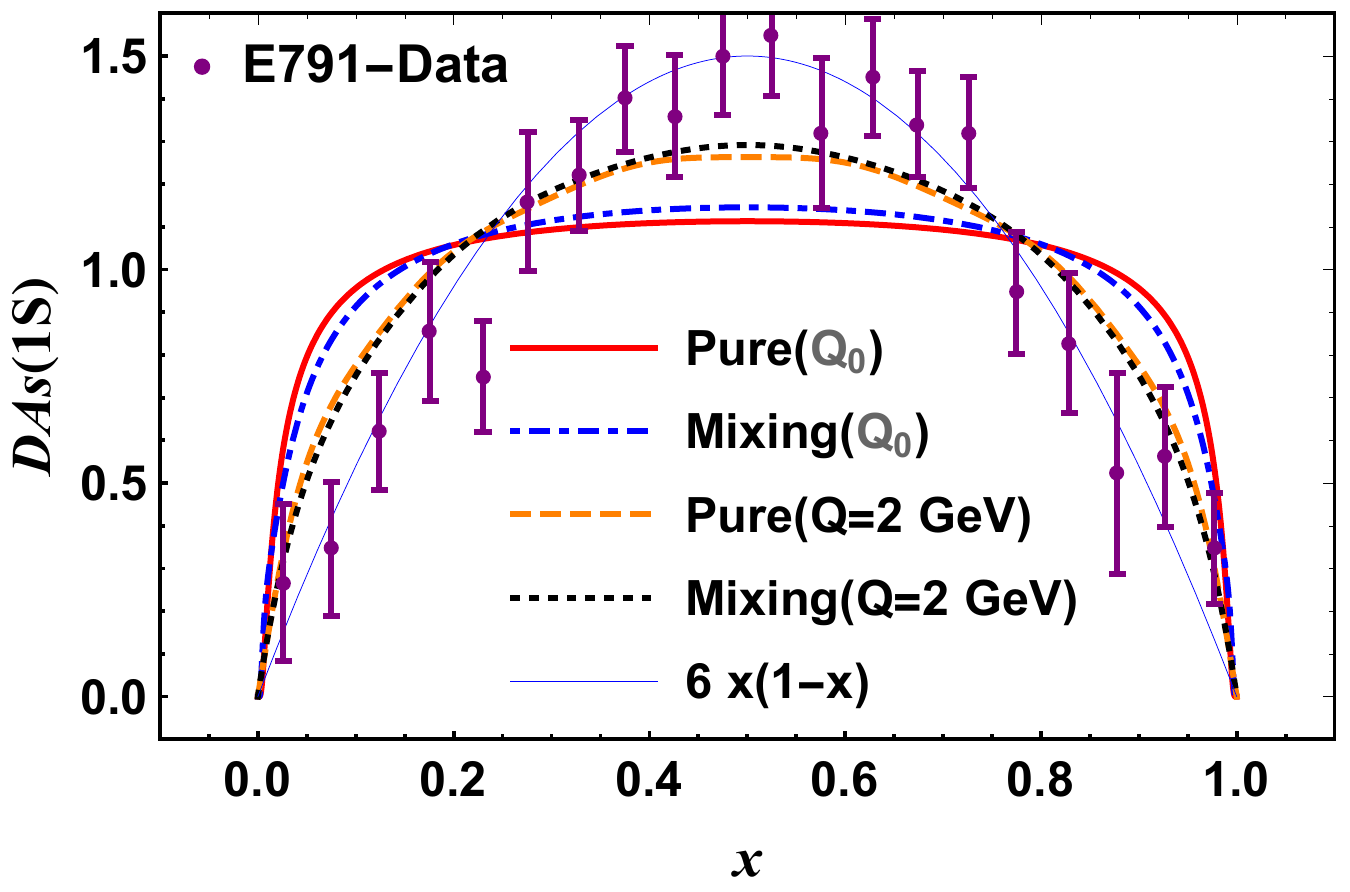}
 \put(17,55){\small (a)}
\end{overpic}
\hfill
\begin{overpic}[width=0.48\textwidth]{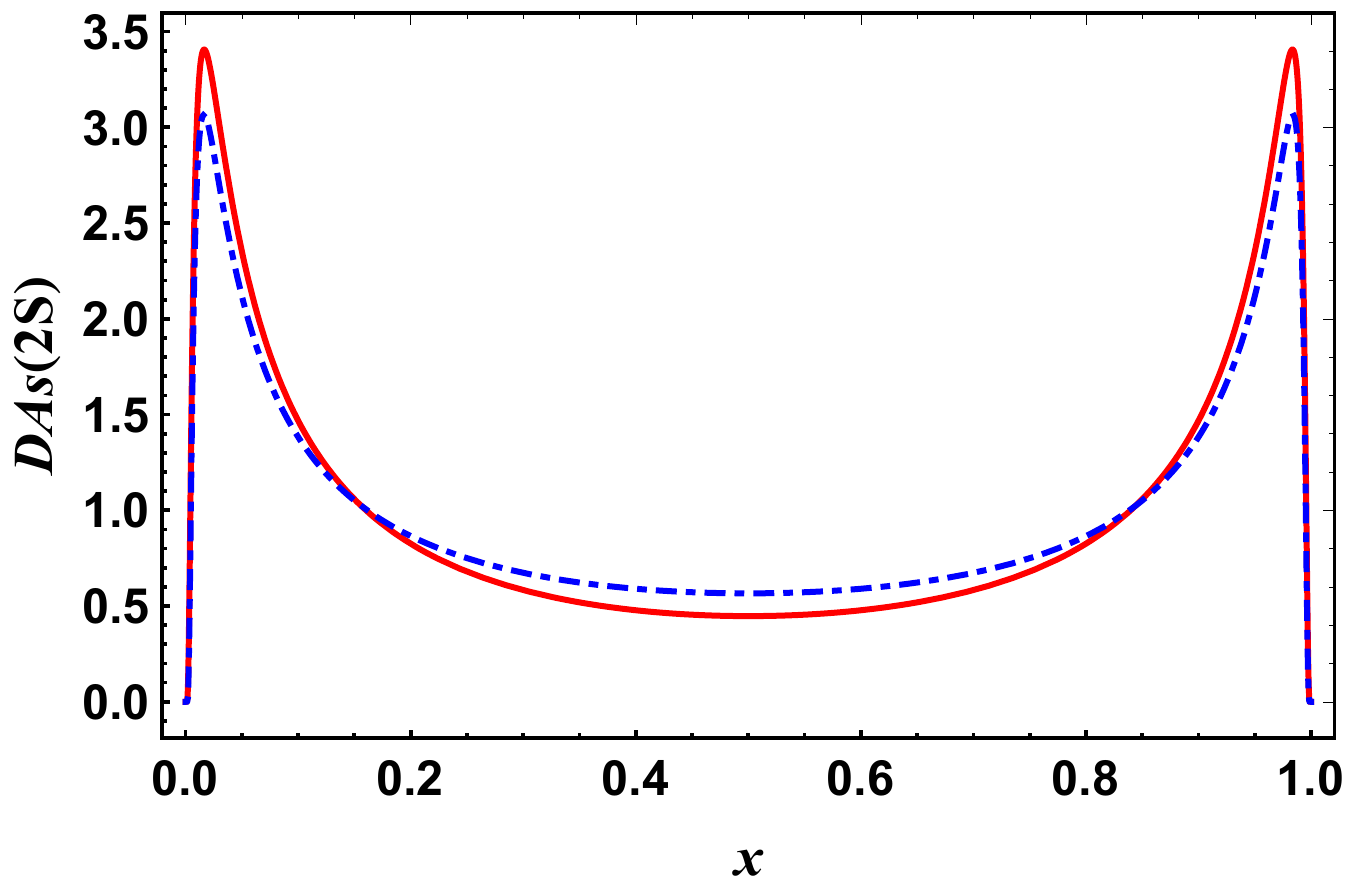}
\put(17,58){\small (b)}
\end{overpic}
\hfill
\begin{overpic}[width=0.48\textwidth]{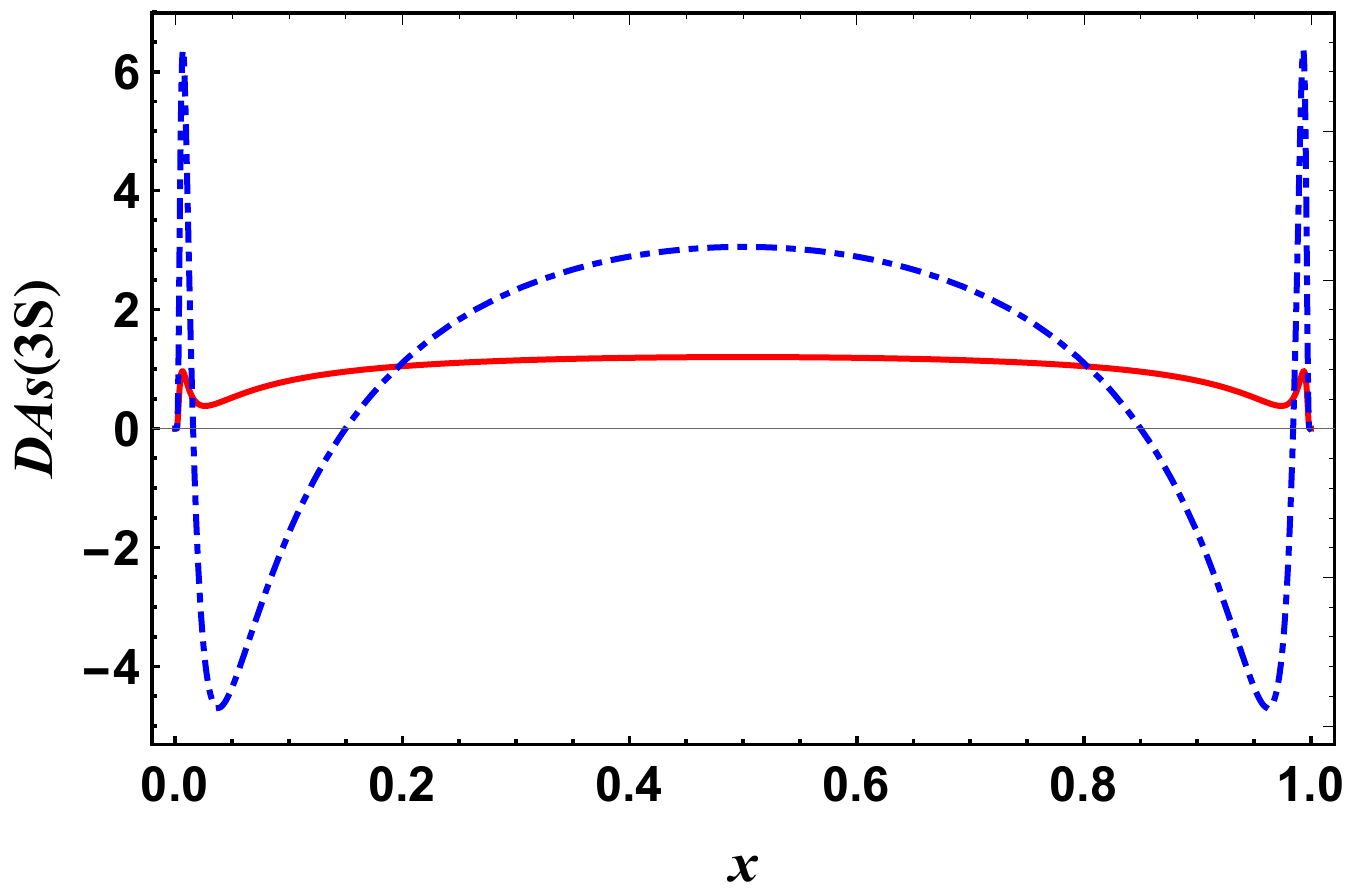}
\put(17,58){\small (c)}
\end{overpic}
\caption{(Color online) DAs of radially excited pions.
(a) DA of the ground-state pion ($\pi (1S)$) at the initial scale $Q_{0}$ and at $Q=2$ GeV for pure and mixed states, compared with the asymptotic form $6x(1-x)$ and E791 data~\cite{E791:2000xcx}; (b) DA of the first excited state ($\pi (2S)$) at $Q_{0}$ for pure and mixed states;
(c) DA of the second excited state ($\pi (3S)$) at $Q_{0}$ for pure and mixed states.}
\label{fig2} 
\end{figure}
%%%%%%%%%%
 For both the pure and mixed states, we observe this trend, i.e., decay constant for $f_{1S}>f_{2S,3S}$ for the realistic values of the harmonic parameter (see Table~\eqref{T1}). However, we see that for the mixed states, the ordering $f_{1S}>f_{2S}>f_{3S}$ is strictly followed for the whole range of $\beta$, which is not true for the pure states. In Table~\eqref{T2}, we compare our results of decay constant evaluated at the $\beta$ values obtained from the variational analysis (see Table~\eqref{T1}) with other calculations~\cite{Ahmady:2022dfv,Xu:2025cyj,Arndt:1999wx} and PDG \cite{10.1093/ptep/ptac097}.
 
After discussing the decay constant, we shift our attention to a closely related quantity: the pion DA. In Fig.~\eqref{fig2}, we show the variation of the DA as a function of the longitudinal quark momentum fraction $x$ for both pure and mixed states. Specifically, in Fig.~\eqref{fig2}(a), we display the ground state DA at both initial scale $Q_{0}=0.49$ GeV as well as at $Q=2$ GeV, which is then compared with the E791 data~\cite{E791:2000xcx} and the asymptotic DA, $\phi(x)=6 x (1-x)|_{Q \rightarrow \infty}$. The evolution of DAs have been carried out through the leading order ERBL evolution method \cite{Lepage:1980fj,Efremov:1979qk}. There is a common feature among the DAs--the mixed state DAs lie above the corresponding DAs of the pure state for the intermediate longitudinal momentum fraction $x\approx0.2-0.8$. For low or high $x$, the DAs for $1S$ and $2S$ pure states lie above the corresponding mixed ones. The DAs of the excited state $3S$ are more interesting as it has two nodes and correspondingly three regions--the mid-$x$ domain where mixed state DA dominates, intermediate-$x$ where pure state DA dominates, and the very low-$x$ domain dominated by mixed state. In Refs. \cite{Xu:2025cyj,Syahbana:2024hkc}, the DA of $\pi(2S)$ is found to have both positive and negative distributions, whereas in our calculations, it is found to be positive for all $x$. 

\subsection{Parton distribution functions (PDFs)}
The unpolarized quark PDF for the pion can be defined as the bilocal correlator of the quark fields at $(z^{+},\vec{z}_{\perp})=0$ \cite{Puhan:2024xdq} as
\begin{eqnarray}
\Phi_{\rm PDF} (x)&=&\frac{1}{2}\int \frac{dz^{-}}{2(2\pi)} ~e^{i(xP^{+}z^{-})/2}\nonumber\\
&& \langle \Pi (P)|\bar{\vartheta}_{u}(-z/2) \gamma^{+} \vartheta_{u}(z/2)|\Pi (P) \rangle. \label{AD16}
\end{eqnarray}
Substituting the quark field operators in Eq.~\eqref{AD16} we obtain,
\begin{eqnarray}
\Phi_{\rm PDF} (x)&=& \int \frac{d^{2}\vec{k}_{\perp}}{2(2\pi)^{3}} \sum_{\lambda_{1},\lambda_{2},\lambda^{\prime}_{1}}\varPsi^{\pi\ast}_{\lambda^{\prime}_{1}\lambda_{2}}(x,\vec{k}_{\perp})\varPsi^{\pi}_{\lambda_{1}\lambda_{2}}(x,\vec{k}_{\perp})\nonumber\\
&& \frac{1}{2xP^{+}}\bar{u}_{\lambda^{\prime}_{1}}(xP^{+},\vec{k}_{\perp})\gamma^{+}u_{\lambda_{1}}(xP^{+},\vec{k}_{\perp})~.
\label{AD17}
\end{eqnarray}
Separating the helicity and space part of the wavefunction and evaluating the spinor products, Eq.~\eqref{AD17} simplified as,
\begin{eqnarray}
\Phi_{\rm PDF} (x)&=& \int \frac{d^{2}\vec{k}_{\perp}}{4x(1-x)} \sqrt{\frac{\vec{k}_{\perp}^{2}+m^{2}}{x(1-x)}} ~|\varPhi(x,\vec{k}_{\perp})|^{2}~.
\label{AD18}
\end{eqnarray}

\begin{figure}[H]
\centering
\begin{overpic}[width=0.48\textwidth]{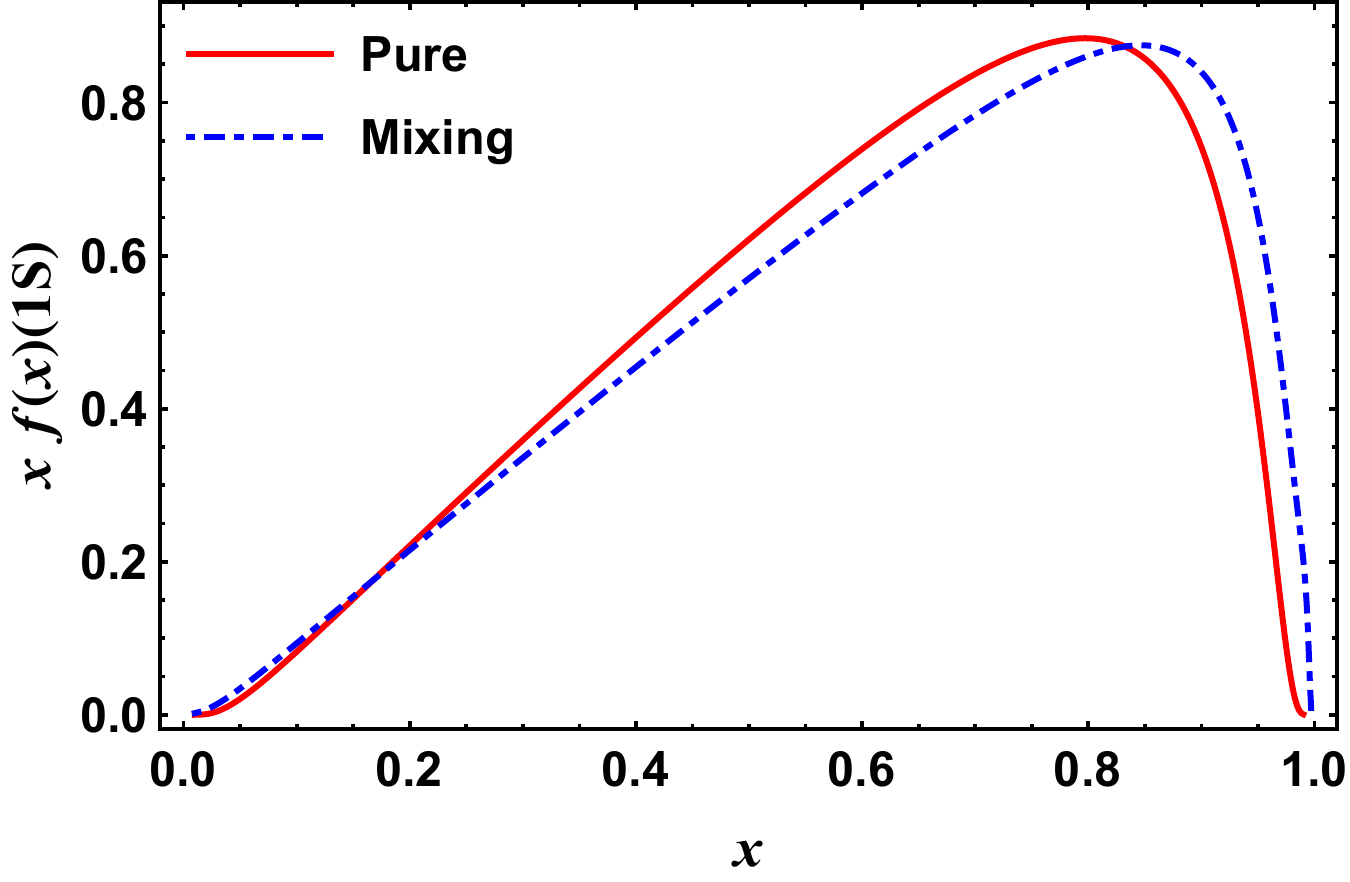}
 \put(50,60){\small (a)}
\end{overpic}
\hfill
\begin{overpic}[width=0.48\textwidth]{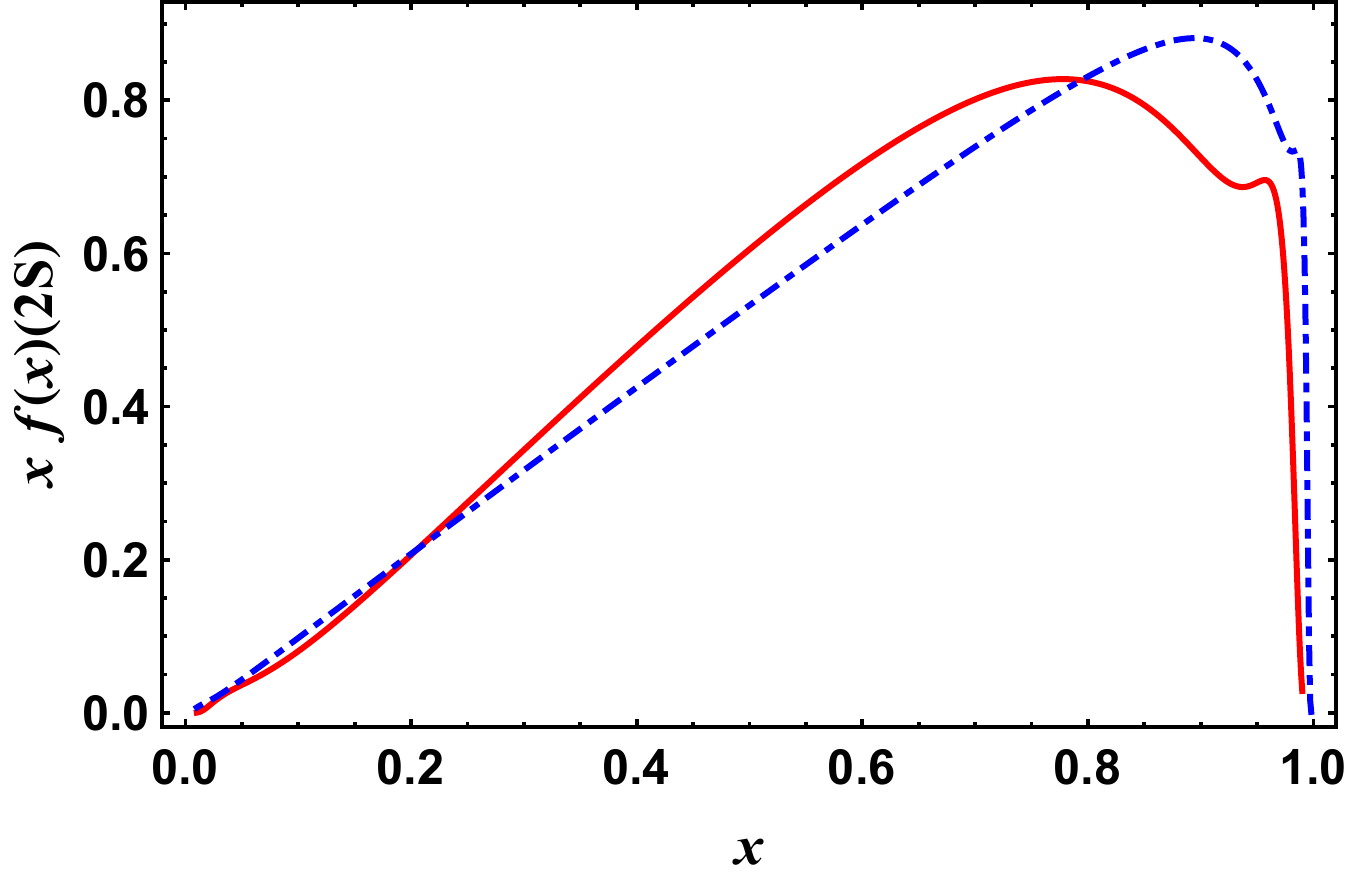}
\put(50,60){\small (b)}
\end{overpic}
\hfill
\begin{overpic}[width=0.48\textwidth]{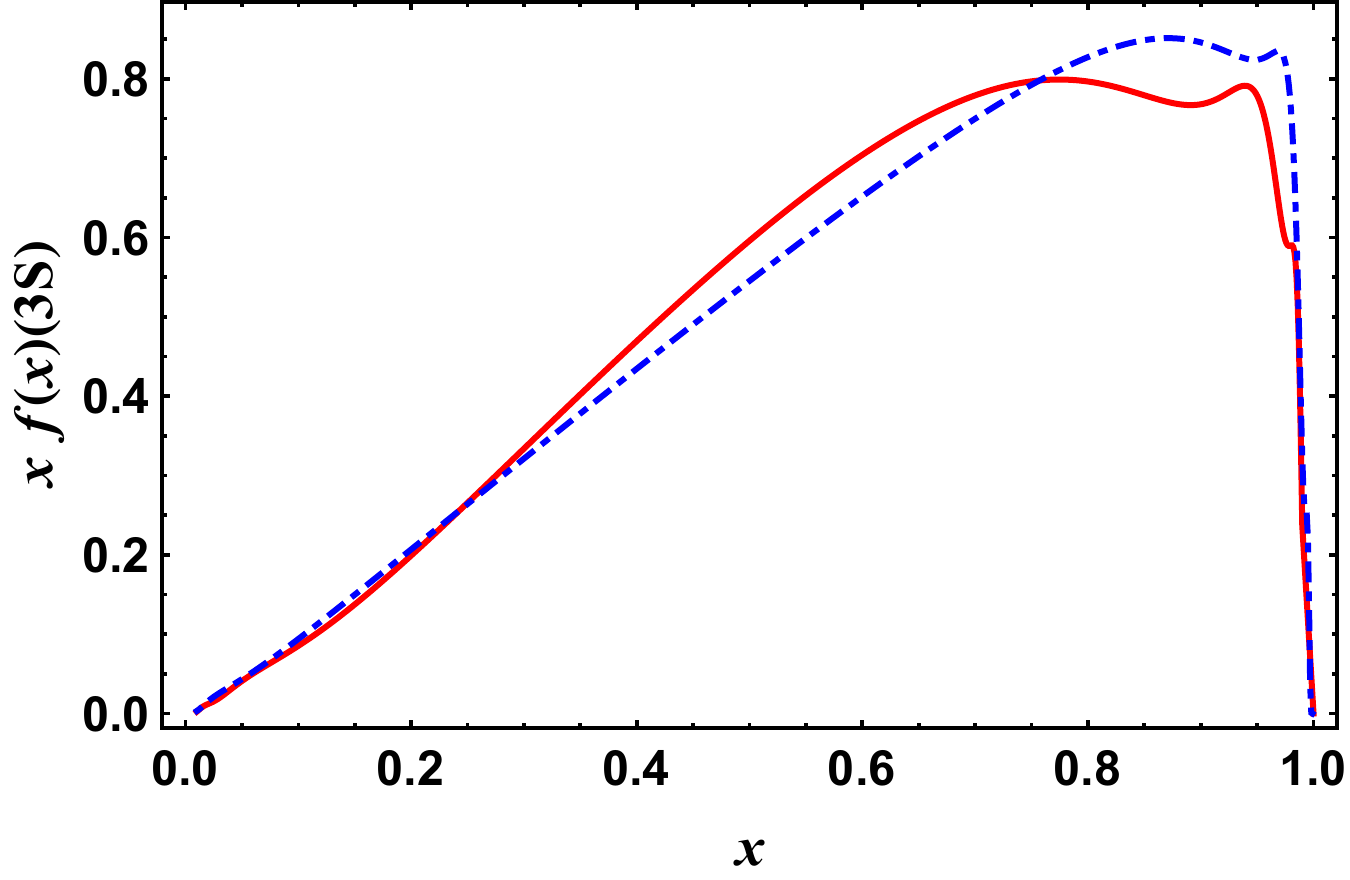}
\put(50,60){\small (c)}
\end{overpic}
\caption{(Color online) PDFs $xf(x)$ of radially excited pions at the model scale $Q^2_{0}=0.20$ GeV$^2$(a) ground state, (b) first excitation, and (c) second excitation, shown for both pure and mixed states.}
\label{fig3} 
\end{figure}

\begin{figure}[H]
\centering
\begin{overpic}[width=0.48\textwidth]{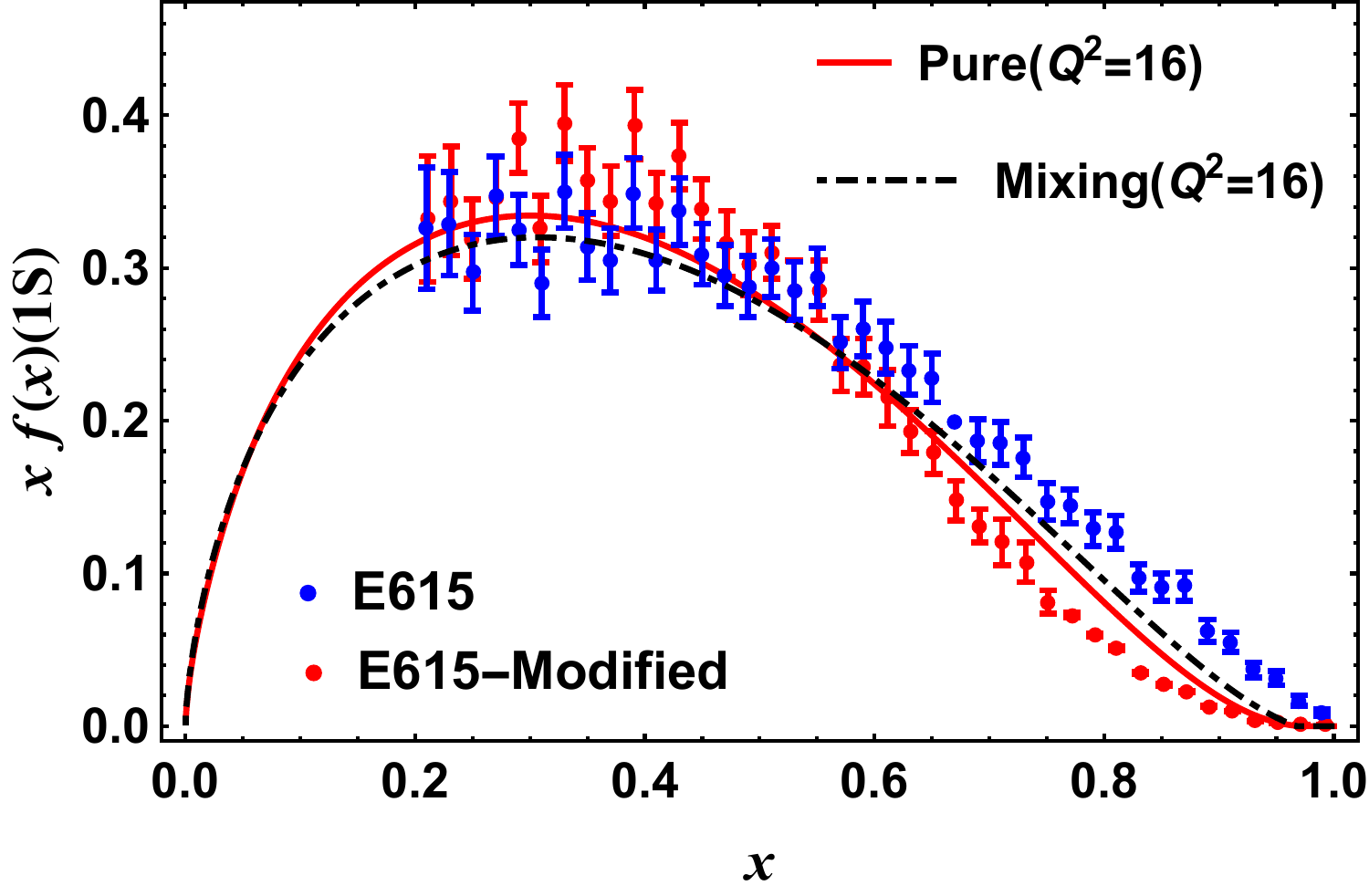}
 \put(50,60){\small (a)}
\end{overpic}
\hfill
\begin{overpic}[width=0.48\textwidth]{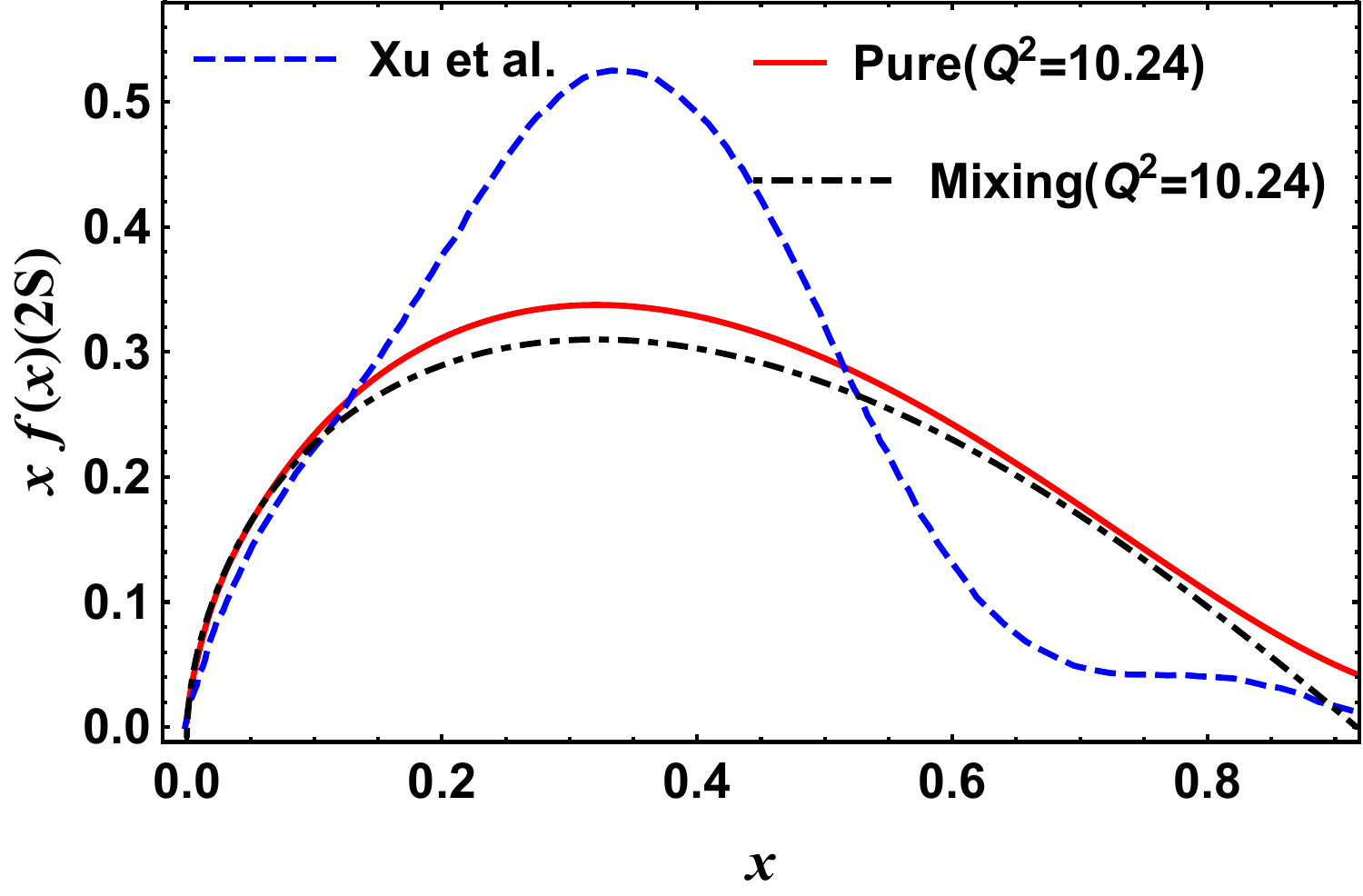}
\put(50,62){\small (b)}
\end{overpic}
\caption{(Color online) PDFs of radially excited pions at different $Q^{2}$: (a) ground state $\pi(1S)$ compared with E615 data~\cite{E615:1989bda, Aicher:2010cb} at $Q^{2}=16$ GeV$^{2}$, and (b) first excitation $\pi(2S)$ compared with Xu et al.~\cite{Xu:2025cyj} at $Q^{2}=10.24$ GeV$^{2}$. Results are shown for both pure and mixed states.}
\label{fig4} 
\end{figure}

In Fig.~\eqref{fig3} (a), (b) and (c), we show the $u$-quark PDF for the ground ($1S$) and two radially excited states ($2S,3S$) of pion, respectively. We observe that for the mixed ground and excited states, the overall distribution shifts to the right in comparison to the pure states. This implies that there is a larger probability of finding the valence $u$-quark at large $x$ values compared to the pure states. In the lower $x$ ($<0.2$), the PDFs for pure and mixed states almost merge with minimal changes. Moreover, we observe kinks at high-$x$ for the $2S$ and $3S$ states in Fig.~\eqref{fig3} (b) and (c). The quark PDFs shown here are at model scale and obey the sum rule $\int f(x)~ dx = 1$.
%%%%%%%%%%%%%%%%%%%%%%%%%%%%%%%%%%%%%%%%%%
Next, we show the $1S$ state $u$-quark PDF which has been evolved to a higher scale $Q^{2}=16$ GeV$^{2}$ through next-to leading order DGLAP evolutions \cite{Miyama:1995bd,Hirai:1997gb,Hirai:1997mm} for comparison with the original~\cite{E615:1989bda} and reanalyzed~\cite{Aicher:2010cb} E615 data in Fig.~\eqref{fig4} (a). Similarly, we display the behavior of PDF for $2S$ mixed and pure states at a scale $Q^{2}=10.24$ GeV$^{2}$ to compare with the prediction of Xu et al.~\cite{Xu:2025cyj}. For the $1S$ state, our prediction is in good agreement with the experimental data obtained for the pion PDF via pion-nucleon collision~\cite{E615:1989bda}. However, the $2S$ state PDF obtained from Eq.~\eqref{AD18} differs from the same obtained from the Bethe-Salpeter amplitude method in Ref.~\cite{Xu:2025cyj}, especially at intermediate values of $x$. For lower values of $x$, the $2S$ state PDF obtained by us is in good agreement with Ref.~\cite{Xu:2025cyj}. We also observed that our results is found to have less magnitude that of lattice QCD predictions \cite{Gao:2021hvs} for pion $2S$ state. 
So far, to the best of our knowledge, PDF measurements of radially excited states have not been performed experimentally. Future experimental facilities such as the EIC~\cite{Accardi:2012qut} may provide access to such data, allowing a more conclusive understanding of the behavior of the distribution functions. In addition, further alternative model calculations and first-principles lattice QCD simulations for the radially excited pion state will be necessary to achieve a more complete and definitive understanding.
%%%%%%%%%%%%%%%%%%%%%%%%%%%%%%
%%%%%%%%%%%%%%%%%%%%%%%%%%%%%%%%%%%%%%%%%%%%%%%%%

\subsection{Electromagnetic form factors (EMFFs)}
The EMFF of the pion is defined by sandwiching the electromagnetic current $J_{\rm em}^{\mu}=\sum_{u,\bar{d}} e_{q} \bar{\vartheta}_{q}\gamma^{\mu}\vartheta_{q}$ between the pion states. The contribution of the $u$ quark to the EMFF is given by,
\begin{eqnarray}
\langle \Pi (P^{\prime})|e_{u}\bar{\vartheta}_{u}(0) \gamma^{+} \vartheta_{u}(0)|\Pi (P) \rangle&=& 2 P^{+} F_{u}(Q^{2}), \label{AD19}
\end{eqnarray}
where the momentum of the initial and final state pion are respectively, $P=(P^{+},\vec{P}_{\perp})$ and $P^{\prime}=(P^{+\prime},\vec{P}^{\prime}_{\perp})$. The momentum transferred in the process $Q=P^{\prime}-P\equiv \Delta$. We choose the Drell-Yan frame $\Delta=(\Delta^{+}=0, \vec{\Delta}_{\perp})$ to evaluate the EMFF. 
Substituting the quark field operators to solve the correlator on the left-hand side of Eq.~\eqref{AD19}, we get,
\begin{eqnarray}
F_{u}(Q^{2})&=& e_{u}\int \frac{dx d^{2}\vec{k}_{\perp}}{2(2\pi)^{3}}\sum_{\lambda_{1},\lambda_{2},\lambda^{\prime}_{1}}\varPsi^{\pi\ast}_{\lambda^{\prime}_{1}\lambda_{2}}(x,\vec{k}^{\prime\prime}_{\perp})\varPsi^{\pi}_{\lambda_{1}\lambda_{2}}(x,\vec{k}^{\prime}_{\perp})\nonumber\\
&& \frac{1}{2xP^{+}}\bar{u}_{\lambda^{\prime}_{1}}(xP^{+},\vec{k}_{\perp}+\frac{\vec{\Delta}_{\perp}}{2})\gamma^{+}u_{\lambda_{1}}(xP^{+},\vec{k}_{\perp}-\frac{\vec{\Delta}_{\perp}}{2}),\nonumber\\\label{AD20}
\end{eqnarray}
where we choose the momenta to be symmetric, i.e., $\vec{k}^{\prime}_{\perp}=\vec{k}_{\perp}-(1-x)\frac{\vec{\Delta}_{\perp}}{2}$ and $\vec{k}^{\prime \prime}_{\perp}=\vec{k}_{\perp}+(1-x)\frac{\vec{\Delta}_{\perp}}{2}$. Summing over the helicity indices, one obtains,
\begin{eqnarray}
F_{u}(Q^{2})&=& e_{u}\int \frac{dx d^{2}\vec{k}_{\perp}}{2(2\pi)^{3}}\sum_{\lambda_{1},\lambda_{2}}\varPsi^{\pi\ast}_{\lambda_{1}\lambda_{2}}(x,\vec{k}^{\prime\prime}_{\perp})\varPsi^{\pi}_{\lambda_{1}\lambda_{2}}(x,\vec{k}^{\prime}_{\perp})\nonumber\\
&=& e_{u}\int \frac{dx d^{2}\vec{k}_{\perp}}{2(2\pi)^{3}} \frac{m^{2}+\vec{k}_{\perp}^{2}+\frac{(1-x)^{2}}{4}\vec{\Delta}_{\perp}^{2}}{\sqrt{(m^{2}+\vec{k}_{\perp}^{\prime\prime 2})(m^{2}+\vec{k}_{\perp}^{\prime 2})}}\nonumber\\
&&\psi(x,\vec{k}_{\perp}^{\prime\prime})\psi(x,\vec{k}_{\perp}^{\prime})\nonumber\\
&=& e_{u}\int \frac{dx d^{2}\vec{k}_{\perp}}{4x(1-x)\sqrt{x(1-x)}}\nonumber\\
&&\frac{m^{2}+\vec{k}_{\perp}^{2}+\frac{(1-x)^{2}}{4}\vec{\Delta}_{\perp}^{2}}{(m^{2}+\vec{k}_{\perp}^{\prime\prime 2})^{1/4}(m^{2}+\vec{k}_{\perp}^{\prime 2})^{1/4}}\Phi(x,\vec{k}_{\perp}^{\prime\prime})\Phi(x,\vec{k}_{\perp}^{\prime})~.\nonumber\\ \label{AD21}
\end{eqnarray}
Form factor for the $\bar{d}$ is obtained similarly by sandwiching the electric current of $\bar{d}$ between pion states. The charge radius of the pion is defined by the slope of the total EMFF in $Q^{2}\rightarrow 0$,
\begin{eqnarray}
\langle r_{\pi}^{2}\rangle=-6\frac{\partial F (Q^{2})}{\partial Q^{2}}\vert_{Q^{2}\rightarrow 0}~,\label{AD22}
\end{eqnarray}
where $F(Q^{2})=F_{u}(Q^{2})+F_{\bar{d}}(Q^{2})$. 

\begin{figure}[H]
\centering
\begin{overpic}[width=0.48\textwidth]{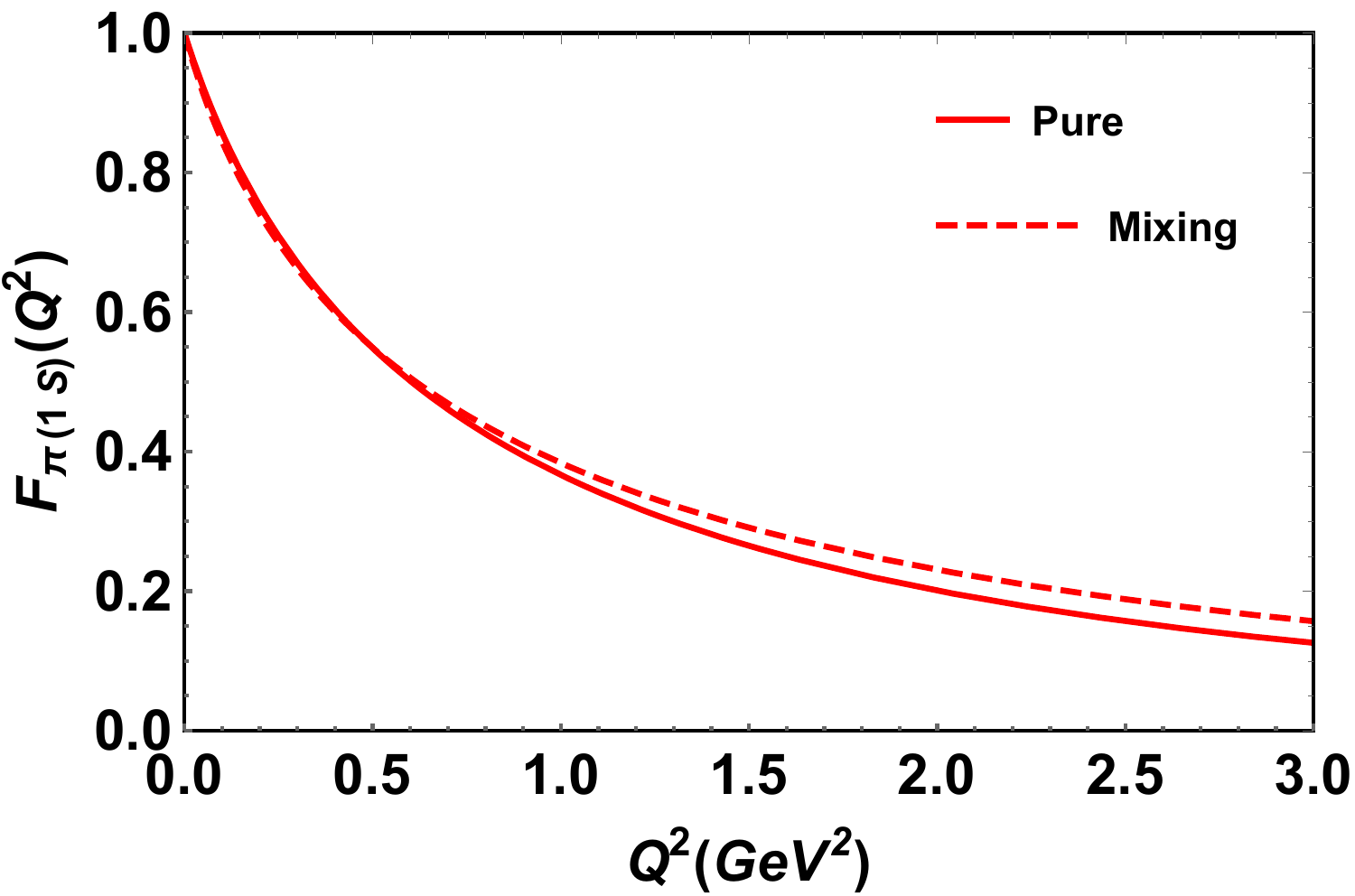}
 \put(50,60){\small (a)}
\end{overpic}
\hfill
\begin{overpic}[width=0.48\textwidth]{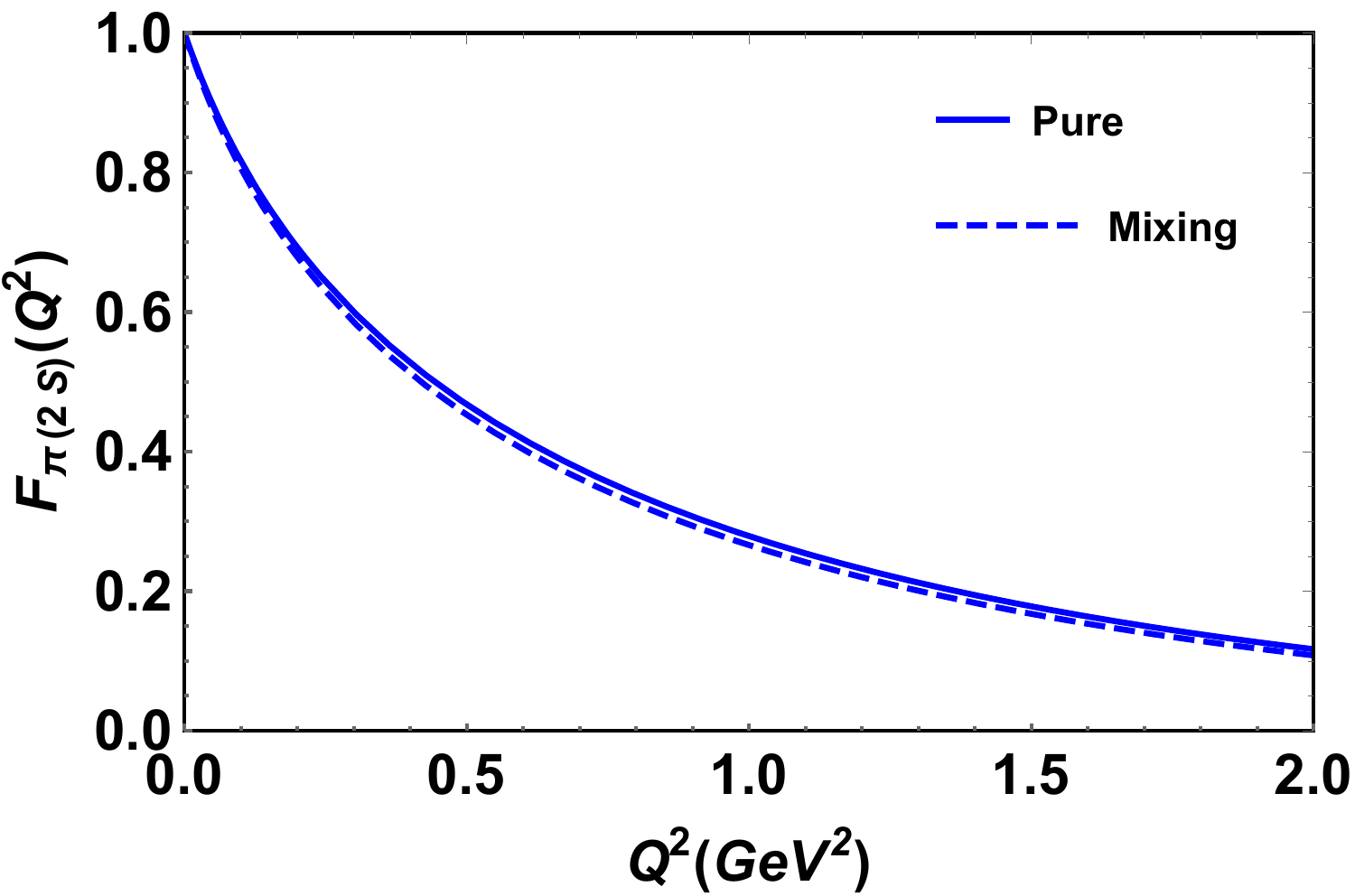}
\put(50,60){\small (b)}
\end{overpic}
\hfill
\begin{overpic}[width=0.48\textwidth]{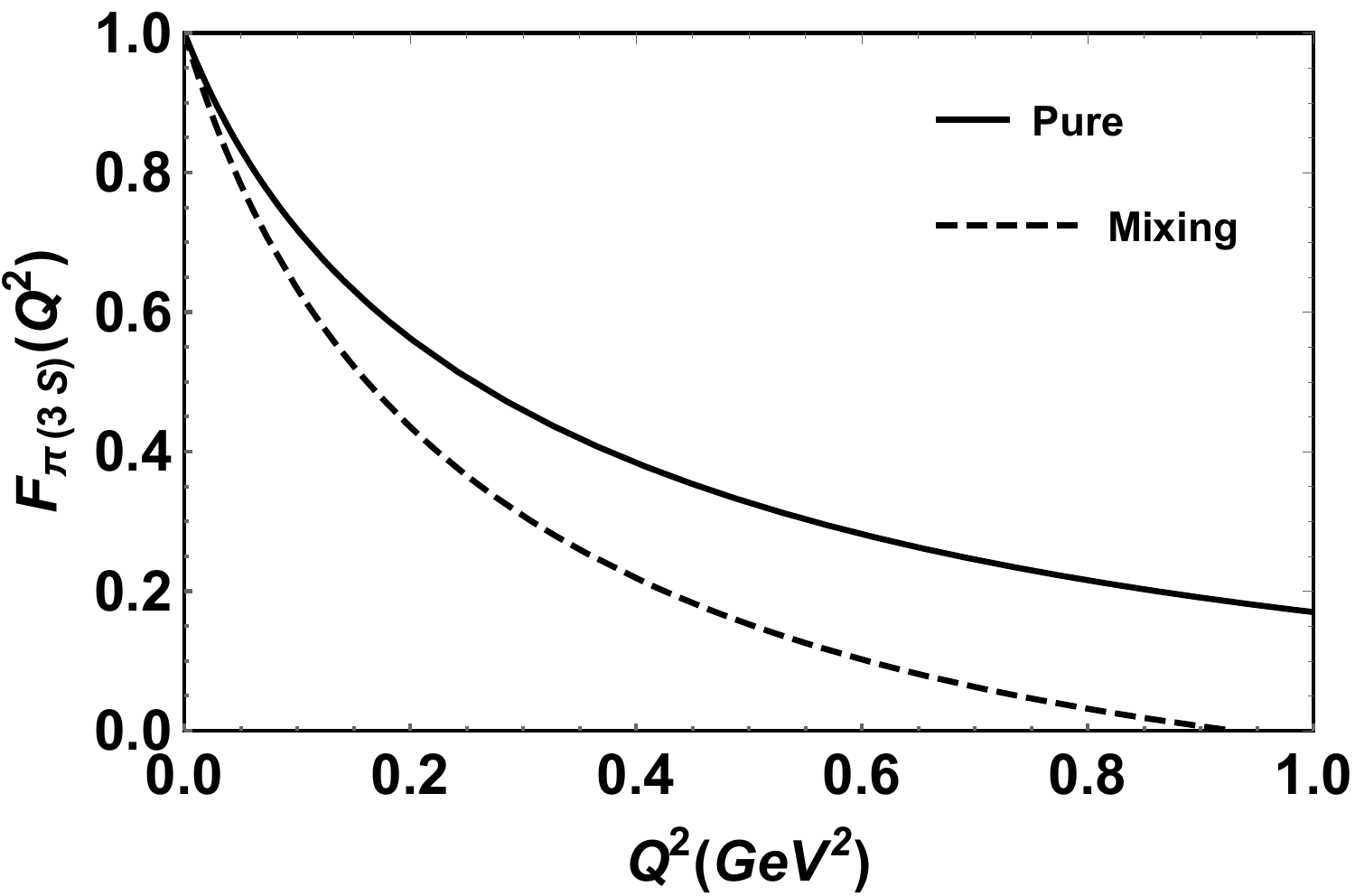}
\put(50,60){\small (c)}
\end{overpic}
\caption{(Color online) EMFFs of radially excited pions at the model scale $Q_{0}$ (a) ground state $\pi(1S)$, (b) first excitation $\pi(2S)$, and (c) second excitation $\pi(3S)$, shown for both pure and mixed states.}
\label{fig5} 
\end{figure}

Equipped with the LFQM formulas for the EMFFs provided in Eq.~\eqref{AD21} and the charge radii in Eq.~\eqref{AD22}, we are now ready to see the variation of the EMFFs as a function of the momentum transferred $Q^{2}$. In Fig.~\eqref{fig5} (a), (b), and (c), we respectively show the EMFFs of $1S$, $2S$, and $3S$ pion states. For the ground state, we see that the EMFF decreases as a function of the momentum transfer in agreement with other theoretical studies (see Ref.~\cite{Puhan:2025pfs} and references therein). For the $1S$ state, we observe that the EMFF corresponding to the mixed state aligns with the pure state for $Q^{2}<1$ GeV$^2$, whereas for high momentum transfer, the mixed state value lies slightly above the pure state. For the $2S$ state, where the EMFFs monotonically decrease with increasing $Q^{2}$, no appreciable change in magnitude is observed between the pure and mixed states. However, our analysis for the second radially excited state $3S$ shows a significant difference between the pure and mixed results. 
\begin{table}[H]
\centering
\begin{tabular}{|c|c|c|c|}
\hline
 $\sqrt{\langle r_{\pi}^{2}\rangle} (fm)$ & $\pi(1S)$ &$\pi(2S)$ & $\pi(3S)$ \\ \hline
 This work (Pure) & 0.45 & 0.74 &  0.98 \\ \hline 
 This work (Mixing) & 0.48 & 0.77 & 1.12  \\ \hline
 Ref.~\cite{Arndt:1999wx} & 0.65 & 0.78 & 0.92 \\ [1ex]
\hline
\end{tabular}
\caption{Charge radii of the pion in fm have been presented and compared with available theoretical predictions \cite{Arndt:1999wx} for all the states.}
\label{T3}
\end{table}
For the $2S$ state, we observe that the value of EMFF lies significantly above the curve denoting the mixed state results in Fig.~\eqref{fig5} (c). 
This implies a significant change in the charge distribution in the impact parameter space, which is essentially the Fourier transform of the EMFF obtained here.

To quantify the charge distribution in the position space, one important parameter is the charge radius defined in Eq.~\eqref{AD22}. We give a comparison of the charge radii of ground and radially excited states for both pure and mixed states in Table~\eqref{T3}. The values tabulated can also be inferred from the plots in Fig.~\eqref{fig5}, where we see almost no change in the slopes for the pure and mixed states curves as $Q^{2}\rightarrow 0$ for $1S$ and $2S$ states, whereas a significant change is observed for $3S$. The charge radius of the ground state pion is found to be less compared to that of $0.657$ of experimental results \cite{NA7:1986vav}. This is due to the larger $\beta$ parameter that we have considered in these calculations.
\section{Summary and conclusions}\label{sum}
In this paper, we have used the light-front constituent quark model to study the structure of the ground and radially excited states of the pion. We determine the shape of the light-front wavefunctions through a variational procedure using an effective QCD Hamiltonian. Two types of trial wavefunctions are considered in the variational procedure. The first consists of harmonic oscillator eigenfunctions, while the second corresponds to mixed wavefunctions formed as orthogonal linear combinations of these eigenfunctions, parametrized by two mixing angles, $\theta$ and $\phi$. In the first case, the variational minimization equations are supplemented by the experimental masses of the low-lying states, $\pi(1S)$ and $\rho(1S)$. These inputs are used to determine the harmonic parameter and to predict the masses of the $\pi(2S)$ and $\pi(3S)$ states. In the second case, the same procedure is followed, but with two additional parameters—the mixing angles—which are chosen to reproduce the PDG masses of the excited states as closely as possible. The consideration of the mixing angles gives us better results of mass spectra with $\chi^{2}=6.51$ MeV, whereas the same without mixing is $84.15$ MeV. After characterization of the shape of the wavefunctions, we have studied the valence-quark structure of the ground and radially excited pion states for both pure and mixed configurations. The distribution amplitudes of the ground and $2S$ states exhibit only weak sensitivity to state mixing, while a significant difference is observed for the $3S$ state. For mixed configuration, the decay constants follow a clear hierarchy, $f_{\pi(1S)}>f_{\pi(2S)}>f_{\pi(3S)}$, which is not maintained for pure states. The parton distribution functions show noticeable differences between pure and mixed states at the model scale, but these differences are largely reduced after evolution to higher momentum scales. The electromagnetic form factors of the $1S$ and $2S$ states are only marginally affected by mixing, whereas pronounced differences appear for the $3S$ state. Finally, the charge radii increase systematically from the ground state to higher radial excitations for both configurations.

This work presents a first systematic investigation of the valence-quark structure of radially excited pion states, highlighting the increasing role of wavefunction mixing for higher excitations. While the present light-front framework successfully captures the sensitivity of key observables to mixing effects, a more complete understanding of excited pion states will require further studies using complementary theoretical approaches, including first-principles lattice QCD, together with improved experimental constraints.

\section{Acknowledgment}
This work was partially supported by the Ministry of Education (MoE), Government of India (A.D.); Board of Research in Nuclear Sciences (BRNS) and Department of Atomic Energy (DAE), Government of India, under Grant No. 57/14/01/2024-BRNS/313 (S.G.); Science and Engineering Research Board (SERB), Anusandhan-National Research Foundation (ANRF), Government of India, under the scheme SERB-POWER Fellowship (Ref No. SPF/2023/000116) (H.D).

\bibliographystyle{elsarticle-num-names} 
\bibliography{references}

\end{document}